\documentclass[nohyper]{JHEP3}

\usepackage[dvips]{graphicx,epsfig}
\usepackage{amsmath}
\usepackage{amsfonts}   % if you want the fonts
\usepackage{amssymb}  
\usepackage{mathrsfs}
\usepackage{exscale}

\def\be{\begin{equation}}
\def\ee{\end{equation}}
\def\bea{\begin{eqnarray}}
\def\eea{\end{eqnarray}}

\def\fun#1#2{\lower3.6pt\vbox{\baselineskip0pt\lineskip.9pt
        \ialign{$\mathsurround=0pt#1\hfill##\hfil$\crcr#2\crcr\sim\crcr}}}
        
\def\expec#1{\langle#1\rangle}

\def\vmu{\mbox{\boldmath${\mu}$}}% Bold-face $\mu$
\def\bfp{\mbox{\bf p}}
\def\bfq{\mbox{\bf q}}

\def\x{{\bf x}}
\def\C{{\bf C}}

\newcommand{\half}{{1\over2}}

\newcommand{\DA}{D\!_A(z)}
\newcommand{\hz}{H(z)}

\newcommand{\Vsur}{V_{\rm survey}}
\newcommand{\Veff}{V_{\rm eff}}

\newcommand{\ihMpc}{h{\rm\;Mpc^{-1}}}
\newcommand{\kmax}{k_{\rm max}}

\title{\mbox{Isocurvature modes and Baryon Acoustic Oscillations II:} gains from combining CMB and Large Scale Structure}

\author{Carmelita Carbone\\  
Dipartimento di Astronomia, Alma Mater Studiorum-Universit\'a di Bologna, via Ranzani 1, I-40127 Bologna, Italy \\
\& 
INAF-Osservatorio Astronomico di Bologna, Via Ranzani 1, I-40127,
Bologna, Italy\\
\&
INFN, Sezione di Bologna, Viale Berti Pichat 6/2, I-40127 Bologna, Italy\\
\email{carmelita.carbone@unibo.it}}
\author{Anna Mangilli \\ ICC-UB (Instituto de Ciencias del Cosmos at Universitad de Barcelona), Marti i Franques 1, Barcelona, Spain\\
\& Institute of Space Sciences (IEEC-CSIC), Fac. Ciencies, Campus UAB, Bellaterra   \\ \email{anna.mangilli@icc.ub.edu}}
\author{ Licia Verde \\ ICC-UB-IEEC (Instituto de Ciencias del Cosmos, Universitad de Barcelona), Mart\'i i Franqu\`es 1, Barcelona, Spain \\
\& ICREA (Instituci\'o Catalana de Rec\`erca i Estudis Avan\c cats) \\ \email{liciaverde@icc.ub.edu}}

\abstract{We consider cosmological parameters estimation in the presence of a non-zero isocurvature contribution in the primordial perturbations. A previous analysis showed that even a tiny amount of isocurvature perturbation, if not accounted for, could  affect standard rulers calibration from  Cosmic Microwave Background observations such as those provided by the Planck mission, affect Baryon Acoustic Oscillations interpretation, and introduce biases in the recovered dark energy properties that are larger than forecasted  statistical errors from future surveys.

Extending  on this work,  here we adopt a general fiducial cosmology
which includes a varying dark energy equation of state parameter and
curvature. Beside Baryon Acoustic Oscillations measurements, we
include the information from the shape of the galaxy power spectrum
and  consider a joint  analysis of a Planck-like Cosmic Microwave
Background probe and a future, space-based,  Large Scale Structure probe not too dissimilar from   recently proposed surveys.

We find that this allows one to break the degeneracies that affect the  Cosmic Microwave Background and Baryon Acoustic Oscillations  combination. As a result, most of
  the cosmological parameter systematic biases arising from an
  incorrect assumption on the isocurvature fraction parameter $f_{iso}$, become negligible with respect
  to the statistical errors.   We find that the Cosmic Microwave
  Background and Large Scale Structure
  combination gives a statistical error $\sigma(f_{iso}) \sim 0.008$,
  even when curvature and a varying dark energy  equation of state are
  included, which is smaller than the error obtained from Cosmic
  Microwave Background alone when flatness and cosmological constant
  are assumed. These results confirm the synergy
  and complementarity  between  Cosmic Microwave Background  and Large
  Scale Structure, and the great potential of   future  and planned
  galaxy surveys.
}

\begin{document}

\section{Introduction}

The standard cosmological model  assumes  adiabatic initial conditions 
for the perturbations. This simple adiabatic picture is well-motivated as it is predicted by the simplest inflationary  single-field models  \cite{mukhanov-adiabatic,Brandenberger92}, and so far it provides an excellent fit to current data  (e.g., \cite{KomatsuWMAP7}).
There is however no a priori reason to discard more general initial conditions involving entropy isocurvature perturbations as they can arise from several different mechanisms e.g., multi-field inflation, \cite{Linde1985,Kofman1986, Mollerach1990,
Polarski1994, Langlois99, Peebles1999, Bartolo2001}, neutrino isocurvature perturbations \cite{Bucher-general-2001}, axion dark matter 
\cite{Axenides:1983hj, Lindeaxion, SeckelTurner, hybrid, TurnerWilczek, LindeLyth, Lyth:1991, Shellard:1997, Kawasaki1995axion} or the curvaton scenario 
\cite{curvaton1, curvaton2, curvaton3}.  
 Pure isocurvature models have been observationally excluded \cite{Efstathiou86, Enqvist02, Page03, Hinshaw06},  but current observations still allow for an admixture of adiabatic and isocurvature contributions \cite{KomatsuWMAP7, Beltran-04, Beltran-05,Valiviita09,Dunkley05,Keskitalo07}. 
 
 To date, Cosmic Microwave Background (CMB) data have been the main source of information not just about cosmological parameters but also about the nature of cosmological initial conditions. 
 Relaxing the assumption of adiabaticity and allowing an isocurvature contribution introduces new degeneracies in the parameters space which weaken considerably  cosmological constraints e.g., \cite{trotta2003, Valiviita09, Kurki05,Langlois00, Valiviita03, Bucher00,Sollom09}.
Up to now,  the CMB temperature power spectrum alone has not been able to  break the degeneracy between the nature of initial  perturbations (i.e. the amount  and properties of an  isocurvature component) and cosmological parameters; adding external data sets somewhat alleviates this issue for some degeneracy directions e.g., \cite{Beltran-04,Dunkley05}.
  As shown in \cite{bucher-pol}, the precision polarization measurements of future and on-going CMB experiments like Planck will be crucial to lift such degeneracies.
 
 Ref.~\cite{Mangilli10} investigated the effect of relaxing the hypothesis of purely adiabatic initial conditions and found that even a tiny isocurvature fraction contribution, if not accounted for, can lead to an incorrect determination of the cosmological parameters and can affect the standard ruler calibration from the CMB. In fact,  the presence of an isocurvature component changes the shape and the location of the CMB acoustic peaks, mimicking the effect of  parameters such as $\Omega_{m_0} h^2$, $H_0$ and $w$ (see  also \cite{Valiviita09} and references therein).
 Ref.~\cite{Mangilli10}  found that this has a crucial effect on
 ``standard rulers", like the sound horizon at radiation drag,
 inferred from CMB observations. This is of relevance for the next
 generation of galaxy surveys which aims at probing with high accuracy
 the late time expansion and thus the nature of dark energy by means
 of Baryon Acoustic Oscillations (BAO) at low redshift ($z<2$). 
In fact, for Large Scale Structure (LSS), even a tiny isocurvature
fraction contribution, if not accounted for,  can affect the  BAO
estimation, introducing systematics in the BAO observables which can be larger than the statistical errors expected from future surveys.

BAO measurements from galaxy surveys have the potential to be an
unprecedented 
powerful probe of the low redshift Universe, and thus of dark energy (\cite{EisensteinHu, SE03, DETF} and references therein).
The BAO in the primordial photon-baryon fluid, responsible for the
characteristic peak structure of the CMB
power spectrum, leave, in fact, an imprint in the large scale matter
distribution and, at each redshift, their physical properties are
related to the size of the sound horizon $r_s(z_d)$ at radiation drag (i.e., when baryons were released from the photons). 
Since the CMB in principle can precisely provide the standard ruler,
i.e. $r_s(z_d)$, by measuring the BAO location at low redshift with
LSS surveys, it is possible to probe the expansion history, i.e., the
Hubble parameter $H(z)$, and the angular diameter distance $D_A(z)$ at
different redshifts, and thus the dark energy properties. 

It has been found that the effect of a number of theoretical
systematics such as non-linearities, bias etc., on the determination of the BAO location can be minimized to the point that BAO are one of the key observables of the next generation dark energy experiments.
However, it is important to investigate and test the robustness of the method to all possible theoretical uncertainties.

The analysis in Ref.~\cite{Mangilli10} focused on the forecasts from a
CMB-only Planck-like experiment to quantify the impact  of the
isocurvature contribution on cosmological parameter estimation and
degeneracies with $r_s(z_d)$ and, therefore, with BAO
measurements. Here we extend the work to a joint analysis of a CMB
Planck-like experiment combined to a future space-based LSS survey
with characteristics not too dissimilar from the proposed Euclid mission\footnote{http://sci.esa.int/euclid}.
One of the key points is to see whether adding the full cosmological
information provided by future LSS surveys could eventually break the
degeneracies introduced by the presence of an isocurvature fraction in
the initial conditions. In the present work in fact, we exploit, in the LSS case, not only the information enclosed in the  BAO positions,
but include the galaxy power spectrum shape adopting the so-called
``$P(k)$-method marginalised over growth information'' \cite{Melita11,Wang06}.
Moreover we consider  spatial curvature and  a more general case of dark energy with a varying equation of state parametrized by $w_0$ and $w_a$. 

The rest of the paper is organized as follows. In Sec.~2 we review
theory and notation for isocurvature perturbations. In Sec.~3  we
outline our methodology, paying particular attention to the method
used to compute expected constraints from LSS surveys, including not
only the BAO location measurements but also the power spectrum shape. 
We present our results in Sec.~4 and the conclusions in Sec.~5.

\section{Isocurvature: theory and notation}\label{sec:notation}

Even if pure isocurvature models have been ruled out \cite{Efstathiou86, Enqvist02, Page03, Hinshaw06}, current observations allow for mixed adiabatic and isocurvature contributions \cite{KomatsuWMAP7, Beltran-04, Beltran-05,Valiviita09,Dunkley05,Keskitalo07}. 

In fact, besides primordial adiabatic perturbations, there can exist
the so-called isocurvature or entropy perturbations. These are
associated to fluctuations in number density between different
components of the cosmological plasma in the early Universe, well
before photon decoupling, and
are generated by stress fluctuations through the causal redistribution of matter under energy-momentum conservation. 

Density perturbations are then produced by non-adiabatic pressure (entropy) perturbations (see \cite{Hu-Sper-Whi97} for a detailed description). 

The entropy perturbation between two particle species (i.e., fluid
components) can be written in terms of the density contrast
$\delta_i=\delta \rho_i/\rho_i$ 
and the equation of state parameter $w_i$ as \footnote{Recall that the continuity equation, which follows from the energy-momentum conservation, takes the form $\dot{\rho} = -3 H (P + \rho)=- 3H\rho (1+w)$.}:
\be
{\cal S}_{XY}= \frac{ \delta_X}{1+ w_X} - \frac{ \delta_Y}{1 + w_Y},
\ee
which quantifies the variation in the particle number densities between two different species, and it is equivalent to ${\cal S}_{XY}= \delta n_X/n_X - \delta n_Y/n_Y$.

The most general description of a primordial perturbation accounts for
5 non-decaying (regular) modes corresponding to each wavenumber: an
adiabatic (AD) growing mode, a baryon isocurvature mode (BI), a cold
dark matter isocurvature mode (CDI), a neutrino density mode (NID) and
a neutrino velocity mode (NIV) (see \cite{Bucher-general-2001} and references therein).
For the dark matter component $\delta n_c/n_c=\delta \rho_c/\rho_c$, so that the CDM isocurvature mode can be written as:
\be
{\cal S}_c \equiv \delta_c - \frac{3}{4} \delta_\gamma,
\ee
where $\delta_X=\delta \rho_X/\rho_X$ is the energy density contrast of the X particle species.  The baryon (b) and the neutrino ($\nu$) isocurvature modes take the form: 
\bea
{\cal S}_b \equiv \delta_b - \frac{3}{4} \delta_\gamma \\
{\cal S}_\nu \equiv \frac{3}{4}  \delta_\nu - \frac{3}{4} \delta_\gamma.
\eea
For the adiabatic mode $S_c=S_\nu=S_b=0$.
	
One of the crucial distinctions between the adiabatic and the
isocurvature models relies on the behavior of the fluctuations at very
early time, during horizon crossing at the epoch of inflation (or an
analogous model for the very early Universe). In the adiabatic case,
constant density perturbations are present initially and imply a
constant curvature on super-horizon scales, while in the pure
isocurvature case there are not initial density fluctuations which,
instead, 
are created from stresses in the radiation-matter component.

An example of isocurvature model is represented by the curvaton scenario  \cite{curvaton1,curvaton2,curvaton3}, which is an alternative to the single field inflationary model.
This model relies on the inflaton, the light scalar field that dominates the background density during inflation, and the curvaton field $\sigma$ which, decaying after inflation, seeds the observable cosmological perturbations. The initial conditions then correspond to purely entropy primordial fluctuations,
namely isocurvature perturbations, because the curvaton field
practically does not contribute to metric perturbations. 
Besides the fact that current data are compatible with this theoretical picture, the curvaton model attracted growing attention because it predicts primordial non-gaussianity features of the local type in the spectrum of primordial perturbations.

\subsection{Notation}

A common parametrization for the isocurvature perturbations is given by \cite{Peiris03}: 
\be\label{e:fiso}
f_{iso}=\frac{\langle {\cal{S} }^2_{rad} \rangle^{1/2}}{\langle {\cal{R}}^2_{rad} \rangle^{1/2}},
\ee
defined as the ratio between the entropy $ {\cal{S}}$ and the
curvature (adiabatic) $ {\cal{R}}$ perturbations evaluated during the
radiation epoch and at a pivot scale $k_0$.  In our case  we set
$0.002 \; {\rm Mpc^{-1}}$, as often done in the literature \cite{WMAP5,KomatsuWMAP7}. 
The correlation coefficient can be then defined in terms of an angle, $\Delta_{k_0}$, such that:
\be
\cos \Delta_{k_0} =\frac{\langle  {\cal{R}}_{rad} {\cal{S} }_{rad} \rangle}{\langle {\cal{R}}^2_{rad} \rangle^{1/2} \langle {\cal{S}}^2_{rad} \rangle^{1/2}}.
\ee
Throughout this paper we will use a sign convention such that $f_{iso}<0$ will correspond to correlated modes.

\subsection{Isocurvature and CMB}

In general the two-point correlation function or power spectra for the
adiabatic mode, the isocurvature mode and their cross-correlation can
be described by two amplitudes, one correlation angle and three
independent spectral indices ($n_{ad}$, $n_{iso}$, $n_{cor}$), so
that, in the case of the CMB, the respective algular power spectra are:
\be\label{eq:Cls-ad}
C_\ell^{ad}=\int \frac{dk}{k}[\Theta^{ad}_\ell(k)]^2 \left( \frac{k}{k_0} \right)^{n_{ad}-1},
\ee
\be\label{eq:Cls-iso}
C_\ell^{iso}=\int \frac{dk}{k}[\Theta^{iso}_\ell(k)]^2 \left( \frac{k}{k_0} \right)^{n_{iso}-1}
\ee
and
\be\label{eq:Cls-corr}
C_\ell^{cor}=\int \frac{dk}{k} \Theta^{ad}_\ell(k)  \Theta^{iso}_\ell(k) \left( \frac{k}{k_0} \right)^{n_{cor}+\frac{1}{2}( n_{ad}+n_{iso})-1}.
\ee
Here $\Theta^{ad}_\ell(k)$ and $ \Theta^{iso}_\ell(k)$ are the radiation transfer functions for adiabatic and isocurvature perturbations that describe how an initial perturbation evolved to a temperature or polarization anisotropy multipole $\ell$.
The total angular power spectrum takes then the form:
\be\label{eq:Cls-tot}
C_\ell=\langle {\cal{R} }^2_{rad} \rangle [C_\ell^{ad}+f^2_{iso} C_\ell^{iso} + 2 f_{iso} \cos \Delta_{k_0} C_\ell^{cor}].
\ee
From Eq.~(\ref{eq:Cls-tot})
 it is clear that, for small isocurvature fractions, the main isocurvature contribution come from the mixing term coefficient ($2 f_{iso} \cos \Delta_{k_0})$.

\subsection{Isocurvature and LSS}\label{sub:isolss}

The power spectra of both
adiabatic and isocurvature perturbations, as well as their
cross-correlation, can be parametrized by three power laws with three amplitudes and three spectral indices $n_{ad}$, $n_{iso}$ and $n_{cor}$:

\bea
\label{primordial_P}
\Delta_{{\cal R}}^2 (k)  &\equiv& 
\frac{k^3}{2\pi^2}\langle{\cal R}_{{\rm rad}}^2 \rangle = 
\frac{k_0^3}{2\pi ^2}
A^2\left(\frac{k}{ k_0}\right)^{n_{ad}-1}\,,\\[2mm] \nonumber
\Delta_{{\cal S}}^2 (k)\!&\equiv&\!
\frac{k^3}{2 \pi^2} \langle {\cal S}_{\rm{rad}}^2 \rangle = \frac{k_0^3}{2\pi ^2}
B^2\left(\frac{k}{k_0}\right)^{n_{iso}-1}\,,\\[2mm]
\Delta_{{\cal R} {\cal S}}^2(k)\!&\equiv&\!\frac{k^3}{2\pi^2}
\nonumber
\langle{\cal R}_{\rm rad}{\cal S}_{\rm rad}\rangle \, \nonumber\\[2mm]
&=& \frac{k_0^3}{2\pi^2} A\,B\,\cos\Delta_{k_0} 
\left(\frac{k}{ k_0}\right)^{n_{cor}+\half(n_{ad}+n_{iso})-1}\,.\nonumber
\eea
As before, ${\cal R}_{\rm rad}$ stands for the curvature perturbation, and
${\cal S}_{\rm rad}=(\delta_{\rm c}-3 \delta_{\gamma} / 4)_{\rm rad}$ for the CDI
perturbation evaluated during the radiation epoch at a pivot scale
$k_0$ such that: $A=\langle{\cal R}_{\rm rad}^2\rangle^{1/2}$ and
$B=\langle{\cal S}_{\rm rad}^2\rangle^{1/2}$. 

The last of Eqs.~(\ref{primordial_P}) refers to the extra-correlation that can be generated by the
partial conversion of isocurvature into adiabatic perturbations after the end of inflation.

Since we are interested in the effect of a model with a mixed
adiabatic+isocurvature CDI  contribution to the LSS, it is useful to give the explicit formula of the shape of the
linear matter power spectrum (e.g.~\cite{Beltran-05}):
\begin{eqnarray}\label{eq:pkiso}
P(k) =  (A^2+B^2)[(1-\alpha) \, P^{\rm ad}(k)+\alpha \, P^{\rm iso}(k)+2\beta\sqrt{\alpha(1-\alpha)} \, P^{\rm cor}(k)]\,.
\end{eqnarray}
Here $P^{\rm ad}$ and $P^{\rm iso}$ are computed from the initial
conditions $({\cal R}_{\rm rad}(k),{\cal S}_{\rm rad}(k)) =((1,0)$, $(0,1))$, and:
\be\label{eq:alphabeta}
\alpha=\frac{f^2_{iso}}{(1+f^2_{iso})}, \; \beta=\cos(\Delta_{k_0}),
\ee
where maximally correlated (anticorrelated) modes correspond to $\beta=+1$ ( $\beta=-1$). 

The extra cross-correlated term is
given by
\begin{equation}
P^{\rm cor}(k) = - (k/k_0)^{n_{cor}} [ P^{\rm ad}(k) P^{\rm iso}(k) ]^{\half}.
\end{equation}

 In this paper we adopt the curvaton scenario as a working example for a model that  gives rise to a small fraction of  correlated CDI isocurvature. As shown in the next sections, our analysis can be generalized to models with an arbitrary amount and type of isocurvature. 
In general, by tuning the decay dynamics of $\sigma$, the curvaton scenario allows for mixed adiabatic and isocurvature fluctuations, 
with any residual isocurvature perturbation correlated or anti-correlated to the adiabatic density one, and with the same tilt for both spectra.
Our findings are therefore quantitative only for this case, however our conclusions will not be too dissimilar for a model such as the axion-like
isocurvature, where $n_{iso}$ is fixed to be 1, given that $n_{ad}$ is not too far away from scale invariance.

In our model we assume $\beta=-1$, $n_s \equiv n_{ad}=n_{iso}$ and $n_{cor}=0$.

\section{Method}\label{sec:method}

In order to extend the analysis done in Ref.~\cite{Mangilli10} on the
impact of an isocurvature contribution on cosmological parameter estimation, 
we compute joint Fisher matrix forecasts \cite{Fisher1935} for a CMB Planck-like experiment \footnote{http://www.sciops.esa.int/index.php?project=PLANCK}
combined with a future space-based LSS survey
with characteristics not too dissimilar from those of the proposed
Euclid mission. In what follow we will refer to such a
survey  as ``Euclid-like'' or ``LSS survey" and use the specifications reported in
\cite{YB:0912.0914}. We adopt the empirical redshift
distribution of H$\alpha$ emission line galaxies derived by
\cite{Geach10} from observed H$\alpha$ luminosity functions,
and the bias function derived by \cite{Orsi10}
using a galaxy formation simulation. In particular, we choose
a flux limit of 3$\times 10^{-16}$erg$\,$s$^{-1}$cm$^{-1}$, a survey
area of 20,000 deg$^2$, a redshift success rate $e=0.35$, a redshift
accuracy of $\sigma_z/(1+z)\le 0.001$, and a redshift range $0.45\leq
z \leq 2.05$. See Tab.~\ref{tab:expsettingPE} for a summary of the experimental specifications.

In both cases we assume a curvaton fiducial model with mixed adiabatic and isocurvature initial conditions. The fiducial value chosen for the isocurvature parameter is 
\mbox{$f_{iso}=-0.08$}.
This is well within the {\footnotesize WMAP7}-only and slightly outside the
{\footnotesize WMAP7+BAO+SN} 95\%CL limits \cite{KomatsuWMAP7,Larson11}. 
On the other hand, these bounds are obtained assuming a $\Lambda$CDM
model, while accounting for curvature and a varying dark energy equation
of state weakens considerably the constraints on the isocurvature contribution \cite{Valiviita09, Dunkley05}.

In any case it is important to stress that the Fisher analysis
results, as found also by Ref.~\cite{Mangilli10}, 
are not strongly affected by the choice of the fiducial values of the isocurvature amount within these limits.

The set of parameters  we use is: 
\be\label{e:parms}
{\bf q_\alpha}= \{\Omega_{m_0}, \Omega_{X_0}, \Omega_{b_0}, h, n_s,
w_0,  w_a,  \log (10^{10} A_{\rm tot}) , f_{iso}\}
\ee
with fiducial values:
\be\label{e:fid}
{\bf \hat{q}_\alpha}= \{ 0.2707 , 0.7293,  0.0451,    0.703 ,   0.966,  -0.95,  0,  1.3909, -0.08 \},
\ee
where $ \Omega_{m_0} $,  $\Omega_{X_0}$  and  $\Omega_{b_0}$  are,
respectively, the total matter, dark energy and baryon density
parameters at present time, $h$ is related to the Hubble parameter by  
$H_0=100 \, h$ Km/(sec Mpc), $n_s$ and $A_{\rm tot}$ are
the scalar spectral index and the total amplitude accounting for both
the adiabatic and the
isocurvature contributions, $w_0$ and $w_a$ are related
to the dark energy equation of state parameter $w(z)$ via the standard CPL parametrization \cite{CPL01,CPL03}:
\be
w(z)=w_0 + w_a \frac{z}{1+z}.
\ee
Finally, $f_{iso}$ is the isocurvature fraction parameter of Eq.~(\ref{e:fiso}) (see \S \ref{sec:notation} for detals).
We assume flatness in the fiducial model but $ \Omega_{m_0}$ and $ \Omega_{X_0}$ are free parameters in the Fisher matrix computation: 
the curvature $\Omega_k=1-  \Omega_{m_0} - \Omega_{X_0}$ is not fixed and it is a parameter in the analysis.

 Recall that, in the case of a CMB-only Planck-like experiment, the
 constraints on the isocurvature contribution are considerably
 weakened when $\Omega_k$ is a  free parameter,  and, when also a
 varying dark energy equation of state is considered, the  cosmological constraints from CMB alone weaken even more  because of degeneracies \cite{Dunkley05}. 

As previously mentioned, in Ref.~\cite{Mangilli10} it was shown that even a small isocurvature contribution, if not accounted for, 
can lead to an incorrect determination of the cosmological parameters and can affect 
the $r_s(z_d)$ calibration from the CMB; this affects the BAO location
estimation and, therefore, leads to systematic shifts in the derived
cosmological  parameters, shifts that result to be larger than
statistical errors from forthcoming LSS surveys.
One of the main purpose of this work  is to see if  this is still the
case for a joint analysis of CMB and LSS when the information from the
$P(k)$-shape are included.  Therefore, here we extend the work of
Ref.~\cite{Mangilli10} not only by
including curvature and dark energy parameters in the analysis, but
also by exploiting BAO location measurements combined to the galaxy power spectrum shape. The additional  cosmological information in the  power spectrum should help break  degeneracies and  thus  indirectly improve constraints on the isocurvature parameter.

%%%%%%%%%%%%%%%%%%%%%%%%%%%%
\subsection{Fisher matrix approach}
%%%%%%%%%%%%%%%%%%%%%%%%%%%%%

We combine the cosmological information of a CMB experiment like
Planck  (enclosed in the $F^{CMB-Planck}$ Fisher matrix) and a galaxy
survey like Euclid (enclosed in the $F^{LSS}$ Fisher matrix), so our full Fisher matrix is:
\be 
F_{ij} \equiv F^{CMB-Planck}_{ij} + F^{LSS}_{ij}.
\label{eq:FisherPE}
\ee

The Fisher matrix is defined as the
second derivative of the natural logarithm of the likelihood surface about the maximum. 
In the approximation that the posterior distribution for the parameters is a
multivariate Gaussian\footnote{In practice,
it can happen that the choice of parametrization                                         
makes the posterior distribution slightly non-Gaussian. However,
for the parametrization chosen here, the error introduced by assuming Gaussianity in the
posterior distribution can be considered as reasonably small, and therefore the
Fisher matrix approach still holds as an excellent approximation for
parameter forecasts.}
with mean $\vmu\equiv\expec{\x}$ and covariance matrix
$\C\equiv\expec{\x\x^t}-\vmu\vmu^t$, 
its elements are given by \cite{VS96,Tegmark,Jungman,Fisher1935}
\begin{align}
\label{eq:fish}
F_{ij} = \frac{1}{2}{\rm Tr}
\left[\C^{-1}{\partial\C\over\partial\theta_i}\C^{-1}{\partial\C\over\partial\theta_j}\right]
+{\partial\vmu\over\partial\theta_i}^t\C^{-1}{\partial\vmu\over\partial\theta_j},
\end{align}
where $\x$ is a N-dimensional vector representing the data set,
whose components $x_i$ are, for example, the 
fluctuations in the galaxy density relative to the mean 
in $N$ disjoint cells that cover the three-dimensional 
survey volume in a fine grid. The $\{\theta_i\}$ denote
the cosmological parameters within the assumed fiducial cosmology.

For the CMB Fisher matrix set up  see appendix \ref{sec:Planck} and
Tab.~\ref{tab:expsettingPE}, for the LSS Fisher matrix set up we
follow closely Ref.~\cite{Melita11} 
as follows.
\begin{table*}
\begin{center}
\begin{tabular}{|cccc|} 
\hline
\hline
Channels (in GHz)&100&143&217\\
Beam FWHM &9.5&7.1&5\\
Temperature sensitivity $\sigma_T$ ($\frac{ \mu K}{K}$) & 2.5 & 2.2 &4.8 \\
Polarization sensitivity $\sigma_P$ ($\frac{\mu K}{K}$) & 4& 4.2 & 9.8\\
%Final noise per arcminute $\mu$K &77.07&57.46&94.42\\
\hline
%\end{center}
%\end{tabular}
%\begin{tabular}{|cccc|} 
\hline
\hline
$z_{min}$& $z_{max}$ & Survey Area (deg$^2$)& $\Delta z_{bin}$\\ %%& $f_{sky}$ \\
% $\bar n$&2.66 $\,10^{-4}$&1.53 $\,10^{-5}$\\
0.45 & 2.05 & 20.000 &0.1 \\
\hline
\end{tabular}
\caption{TOP: experimental specifications for the Planck
  experiment. BOTTOM:  LSS survey settings, i.e. redshift range
 $z_{min}$ and  $z_{max}$, survey area (A) in squared degrees, and width of the redshift bin $\Delta z_{bin}$.    
}
\label{tab:expsettingPE}
\end{center}
\end{table*}

\subsection{Large scale structure treatment}
In order to explore the cosmological parameter
constraints from a given redshift survey, we need to specify
the measurement uncertainties of the galaxy power spectrum.
The statistical error on the measurement of the galaxy
power spectrum $P_{\rm g}(k)$
at a given wave-number bin is \cite{FKP}
\begin{equation}
\left[\frac{\Delta P_{\rm g}}{P_{\rm g}}\right]^2=
\frac{2(2\pi)^2 }{\Vsur k^2\Delta k\Delta \mu}
\left[1+\frac1{n_{\rm g}P_{\rm g}}\right]^2,
\label{eqn:pkerror}
\end{equation}
where $n_{\rm g}$ is the mean number density of galaxies, 
$\Vsur$ is the comoving survey volume of the galaxy survey, and $\mu$
is the cosine of the angle between $\bf{k}$ and the line-of-sight
direction $\mu = \vec{k}\cdot \hat{r}/k$.

In general, the \emph{observed} galaxy power spectrum is different
from the \emph{true} spectrum, and it can be reconstructed approximately
assuming a reference cosmology (which we consider to be our fiducial
cosmology) as (e.g.~\cite{SE03})
\begin{align}
P_{\rm obs}(k_{{\rm ref}\perp},k_{{\rm ref}\parallel},z)
=\frac {\DA _{\rm ref} ^2 \hz}{\DA ^2 \hz _{\rm ref}} P_{\rm g}(k_{{\rm ref}\perp},k_{{\rm ref}\parallel},z)
+P_{\rm shot}\,,
\label{eq:Pobs}
\end{align}
where
\begin{align}
P_{\rm g}(k_{{\rm ref}\perp},k_{{\rm
    ref}\parallel},z)=b(z)^2\left[1+\beta(z) 
\frac{k_{{\rm ref}\parallel}^2}{k_{{\rm ref}\perp}^2+k_{{\rm ref}\parallel}^2}\right]^2\times
P_{{\rm matter}}(k,z)\,.
\label{eq:Pg}
\end{align}
In Eq.~(\ref{eq:Pobs}), $H(z)$ and $D_A(z)$ are the Hubble parameter and the angular
diameter distance, respectively, and the prefactor 
$(\DA _{\rm ref} ^2 \hz)/(\DA ^2 \hz _{\rm ref})$ encapsulates the
geometrical distortions due to a reference cosmology different from
the true one \cite{SE03,9605017}.  The quantities evaluated in the reference cosmology are
distinguished by the subscript `ref', while those in the true cosmology have no
subscript. $k_\perp$ and $k_\parallel$ are the wave-numbers across and along
the line of sight in the true cosmology, and they are related to the 
wave-numbers calculated assuming the reference
cosmology by
$k_{{\rm ref}\perp} = k_\perp D_A(z)/D_A(z)_{\rm ref}$ and
$k_{{\rm ref}\parallel} = k_\parallel H(z)_{\rm ref}/H(z)$. 
$P_{shot}$ is the unknown  shot noise that 
remains even after the conventional shot noise of inverse number density has been 
subtracted  and which we assume to be white noise \cite{SE03}.  Such contribution could arise from galaxy clustering bias even
on large scales due to local bias \cite{Seljak00} or to misestimates of the local density.
In Eq.~(\ref{eq:Pg}), $b(z)$ is the \emph{linear bias} factor between galaxy and
matter density distributions, and $\beta(z)=f_g(z)/b(z)$ is the linear 
redshift-space distortion parameter \cite{Kaiser1987}. Thus our analysis is limited to large enough scales where bias is scale-independent (we will return to this below).
 We estimate $f_g(z)$ by integration of the evolution equations of the
 linear density perturbations. 
%For the
The linear matter power spectrum 
$P_{{\rm matter}}(k,z)$ in Eq.~(\ref{eq:Pobs}) takes the form
\begin{eqnarray}
P_{{\rm matter}}(k,z)=\frac{8\pi^2c^4k_0 A_{\rm tot} }{25
  H_0^4\Omega_{m}^2} T^2(k) \left [\frac{G(z)}{G(z=0)}\right]^2
\left(\frac{k}{k_0}\right)^{n_s}e^{-k^2\mu^2\sigma_r^2},
\label{eq:Pm}
\end{eqnarray}
where $G(z)$ is the linear
growth-factor, whose fiducial value in each redshift bin is computed 
through numerical integration of the differential equations governing the growth
of linear perturbations in the presence of dark energy
\cite{astro-ph/0305286}. 
The linear transfer function $T(k)$ depends on matter and baryon 
densities (neglecting dark energy at early times), 
and is computed in each redshift bin using
CAMB\footnote{http://camb.info/} \cite{CAMB} modified to include the isocurvature contribution.
The parameter $A_{\rm tot}$ refers to the amplitude accounting for both
the adiabatic and the isocurvature contributions as described in \S
\ref{sub:isolss}, and $n_s$ is the spectral index which in our case
reads $n_s \equiv n_{ad} = n_{iso}$, 
so that it can be  factorized out in the $P_{{\rm matter}}(k,z)$
expression in Eq.~(\ref{eq:Pm}).

Moreover, in Eq.~(\ref{eq:Pm}) we have added the damping factor
exp($-k^2\mu^2\sigma_r^2$),
due to redshift uncertainties, where $\sigma_r=(\partial r/\partial
z)\sigma_z$, $r(z)$ being the comoving
distance \cite{0904.2218,SE03}. 

From the above, it should be clear  that, when adding LSS, we exploit the 
information from both the galaxy power spectrum shape and BAO distance
indicators. 

In each redshift shell, with size $\Delta z_{bin}=0.1$ and centred
at redshift $z_i$, we choose the following set of parameters to
describe the observed power spectrum  $P_{\rm obs}(k_{{\rm ref}\perp},k_{{\rm ref}\parallel},z)$: 
\begin{equation}
\left\{H(z_i), D_A(z_i), \bar{G}(z_i), \beta(z_i), 
P_{shot}^i, \omega_m, \omega_b, f_{iso}, n_s, h\right\}, 
\end{equation}
where  $f_{iso}$ is the isocurvature fraction parameter as defined in \S \ref{sec:notation},
$\omega_m=\Omega_{m_0} h^2$, and $\omega_b=\Omega_{b_0} h^2$.
Finally,  since $G(z)$, $b(z)$, and the power spectrum
normalization $P_0$ are completely degenerate, we have introduced the
quantity $\bar{G}(z_i)=(P_0)^{0.5} b(z_i) G(z_i)/G(z_0)$ \cite{0710.3885}.

In the limit where the survey volume is much larger than the scale of 
any features in $P_{\rm obs}(k)$, it has been shown \cite{Tegmark97}
that it is possible to redefine $x_n$ to be not the density fluctuation
in the $n^{th}$ spatial volume element, but the
average power measured with the FKP method \cite{FKP} in a thin shell 
of radius $k_n$ in Fourier space. Under these assumptions,
the redshift survey Fisher matrix can be approximated as \cite{Tegmark,Tegmark97}
\begin{eqnarray}
F_{ij}^{\rm LSS}&=&\int_{\vec{k}_{\rm min}} ^ {\vec{k}_{\rm max}} \frac{\partial
  \ln P_{\rm obs}(\vec{k})}{\partial p_i} \frac{\partial \ln P_{\rm
    obs}(\vec{k})}{\partial p_j} \Veff(\vec{k})
\frac{d\vec{k}}{2(2 \pi)^3}\\ \nonumber
&=&\int_{-1}^{1} \int_{k_{\rm min}}^{\kmax}\frac{\partial \ln
  P_{\rm obs}(k,\mu)}{\partial p_i} \frac{\partial \ln P_{\rm obs}(k,\mu)}{\partial p_j} 
\Veff(k,\mu) \frac{2\pi k^2 dk d\mu}{2(2\pi)^3},
\label{Fisher}                 
\end{eqnarray}
where the derivatives are evaluated at the parameter values $p_i$
of the fiducial model,
and $\Veff$ is the effective volume of the survey:
\begin{eqnarray}
\Veff(k,\mu) =
\left [ \frac{{n_{\rm g}}P_{\rm g}(k,\mu)}{{n_{\rm g}}P_{\rm g}(k,\mu)+1} \right ]^2 \Vsur,
\label{V_eff}                                                                                        
\end{eqnarray}
where we have assumed that the comoving number density
$n_{\rm g}$ is constant in position.
Due to azimuthal symmetry around the line of sight,
the three-dimensional galaxy redshift power spectrum
$P_{\rm obs}(\vec{k})$ depends only on $k$ and $\mu$, i.e. is reduced
to two dimensions by symmetry \cite{SE03}.

To minimize non-linear effects, we restrict wave-numbers to the 
quasi-linear regime, so that $\kmax$ is given
by requiring that the variance of matter fluctuations in a sphere of
radius $R$ is $\sigma^2(R)=0.25$ for $R=\pi/(2\kmax)$. This gives 
$\kmax\simeq 0.1 \ihMpc$ at $z=0$ and $\kmax\simeq 0.2 \ihMpc$
at $z=1$, well within the quasi-linear regime. In addition, we
impose a uniform upper limit of $\kmax \leq 0.2\ihMpc$ (i.e. $\kmax=0.2
\ihMpc$ at $z>1$), to ensure that we are only considering the
conservative linear regime, essentially unaffected by non-linear effects.
In each redshift bin we fix $k_{\rm min}= 10^{-4} h$/Mpc, and we have
verified that changing the survey maximum scale $k_{\rm min}$ with
the shell volume has almost no effect on the results.

We do not include information from the amplitude $\bar{G}(z_i)$ and
the linear redshift space distortions $\beta(z_i)$, so we marginalise over
these parameters and also over $P_{shot}^i$. Moreover, since we are limiting our
analysis to quasi-linear scales, we do not include the effect of the
non-linear (incoherent velocities) redshift space distortions, the
so-called ``Fingers of God'' (FoG). In
fact, as shown in Ref.~\cite{Melita11}, on such scales the effect of
marginalisation over incoherent velocities can be compensated by the
inclusion of growth information, and viceversa, if growth is
marginalised over, the FoG effect is in some sense already taken into
account, due to the tight correlation between these two effects on
linear scales.

We project $\bfp=\{H(z_i), D_A(z_i), \omega_m, \omega_b, f_{iso}, n_s, h\}$
into the final sets $\bfq$ of cosmological parameters described in Eq.~(\ref{e:parms})-(\ref{e:fid})
\cite{Wang06,Wang08a}. 
In this way we adopt the so-called  ``full $P(k)$--method, marginalized
over growth--information'' \cite{1006.3517}, and,
to change from one set of parameters to another, we use \cite{Wang06}
\begin{equation}
F_{\alpha \beta}^{\rm LSS}= \sum_{ij} \frac{\partial p_i}{\partial q_{\alpha}}\,
F_{ij}^{\rm LSS}\, \frac{\partial p_j}{\partial q_{\beta}},
\label{eq:Fisherconv}
\end{equation}
where $F_{\alpha \beta}^{\rm LSS}$ is the survey 
Fisher matrix for the set of parameters $\bfq$, and 
$F_{ij}^{\rm LSS}$ is the survey Fisher matrix for the set of equivalent 
parameters $\bfp$.

We combine the LSS Euclid-like survey with CMB information from the Planck satellite, using the  parameter set as specified in   Appendix \ref{sec:Planck}.

After marginalization
over the optical depth, we propagate the Planck CMB Fisher matrix $F_{ij}^{CMB-Planck}$
into the final sets of parameters ${\bf q}$,
by using the appropriate Jacobian for the involved parameter
transformation.

The 1--$\sigma$ error on $q_\alpha$ marginalized over the
other parameters is
$\sigma(q_\alpha)=\sqrt{({F}^{-1})_{\alpha\alpha}}$, 
where ${F}^{-1}$ is the inverse of the Fisher matrix.
We then consider constraints in a two-parameter
subspace, marginalizing over the remaining parameters, in order 
to study the covariance between $f_{iso}$ and other relevant parameters. 

\subsection{Correlation and shifts}

To quantify the level of degeneracy between
the different parameters, we estimate the so-called
correlation coefficients, given by
\begin{equation}
r\equiv \frac{({F}^{-1})_{p_\alpha p_\beta}}
{\sqrt{({F}^{-1})_{p_\alpha p_\alpha}({F}^{-1})_{p_\beta p_\beta}}},
\label{correlation}
\end{equation}
where $p_\alpha$ denotes one of the model parameters.  When $|r|=1$, the two parameters are totally degenerate, while
$r=0$ means they are uncorrelated.

Furthermore, we use another powerful tool encoded in the Fisher analysis which allows to calculate the shift in the best fit parameters $\delta\theta_\alpha$ in the case that the isocurvature amount parameter $f_{iso}$ is fixed to a wrong fiducial value, without recomputing the full covariance matrix.

In general, given a number $p$ of parameters  $\Psi_\gamma \; (\gamma=1,..,p)$ fixed to an incorrect value, which differs from the ``true" value by an amount $\delta \Psi_\gamma$, the  resulting shift on the  best fit value of the other $n$ parameters $ \theta_\alpha \; (\alpha=1,...,n)$ is (e.g.~\cite{Heavens-Kitching-Verde07}): 
\be
\delta \theta_\alpha = - \left(F^{-1}\right)_{\alpha \beta} \, S_{\beta  \gamma} \delta \Psi_\gamma. %\nonumber
\label{eq:shift} 
\ee
Here $\left(F^{-1}\right)_{\alpha \beta}$ is the sub-matrix of the
inverse Fisher matrix in Eq.~(\ref{eq:FisherPE}),  corresponding to
the $\theta_\alpha$ parameters (i.e. the inverse of the Fisher matrix,
without the rows and columns corresponding to the ``incorrect"
parameters), and $S_{\beta  \gamma}$ is the Fisher sub-matrix including also the $\Psi_\gamma$ parameters.
The shifts in the best fit parameters induced by setting the amount of isocurvature $f_{iso}$ to an incorrect value is:
$\delta \Psi_\gamma=\delta f_{iso}$ where \mbox{$\theta_\alpha=\{
  \Omega_{m_0}, \Omega_{X_0}, \Omega_{b_0}, h, n_s, w_0, w_a, \log
  (10^{10} A_{\rm tot}) \}$}.

\section{Results}
\begin{figure*}%[!ht]
\begin{tabular}{l l}
\centering
\includegraphics[width=7.3cm]{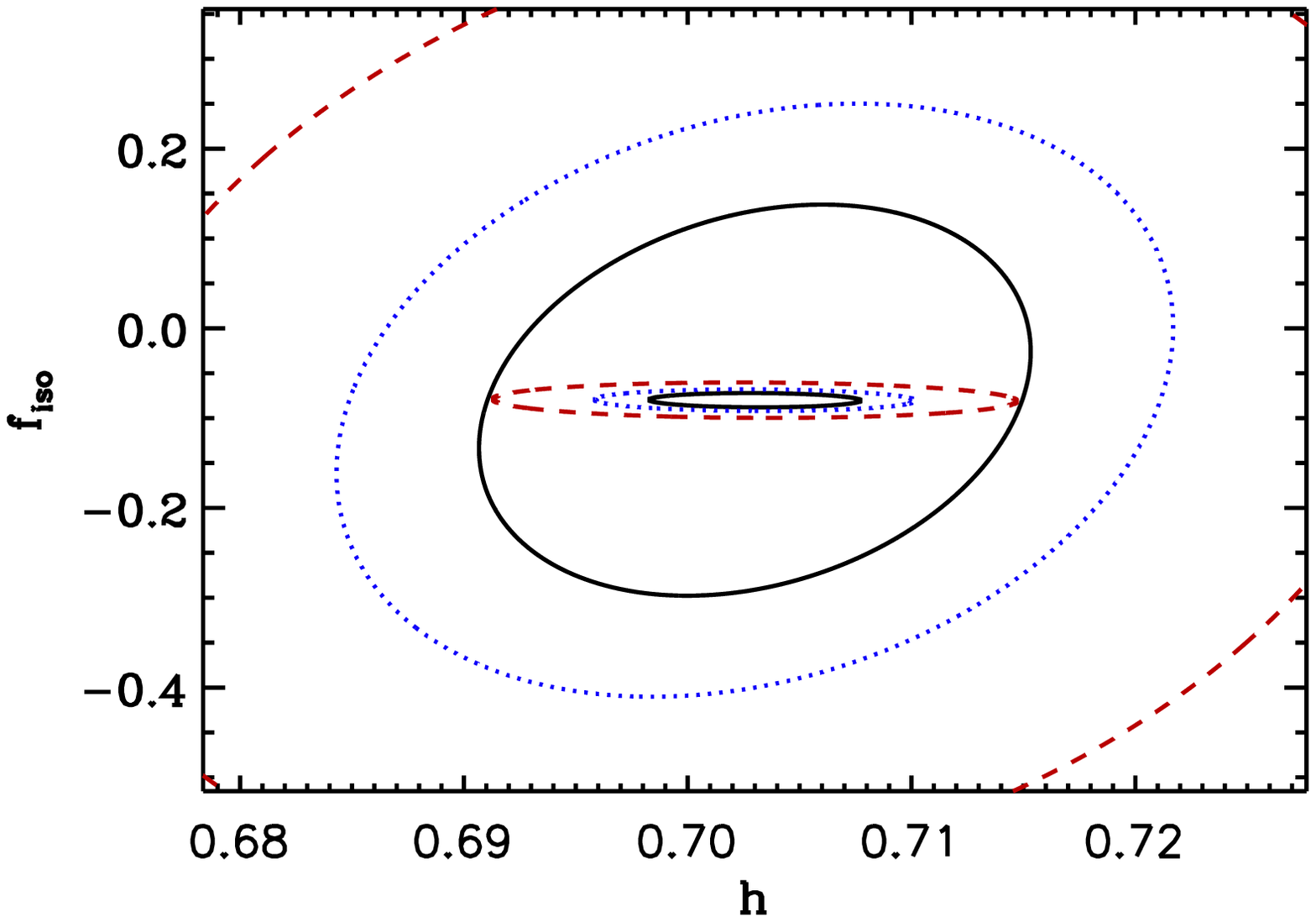}&
\includegraphics[width=7.3cm]{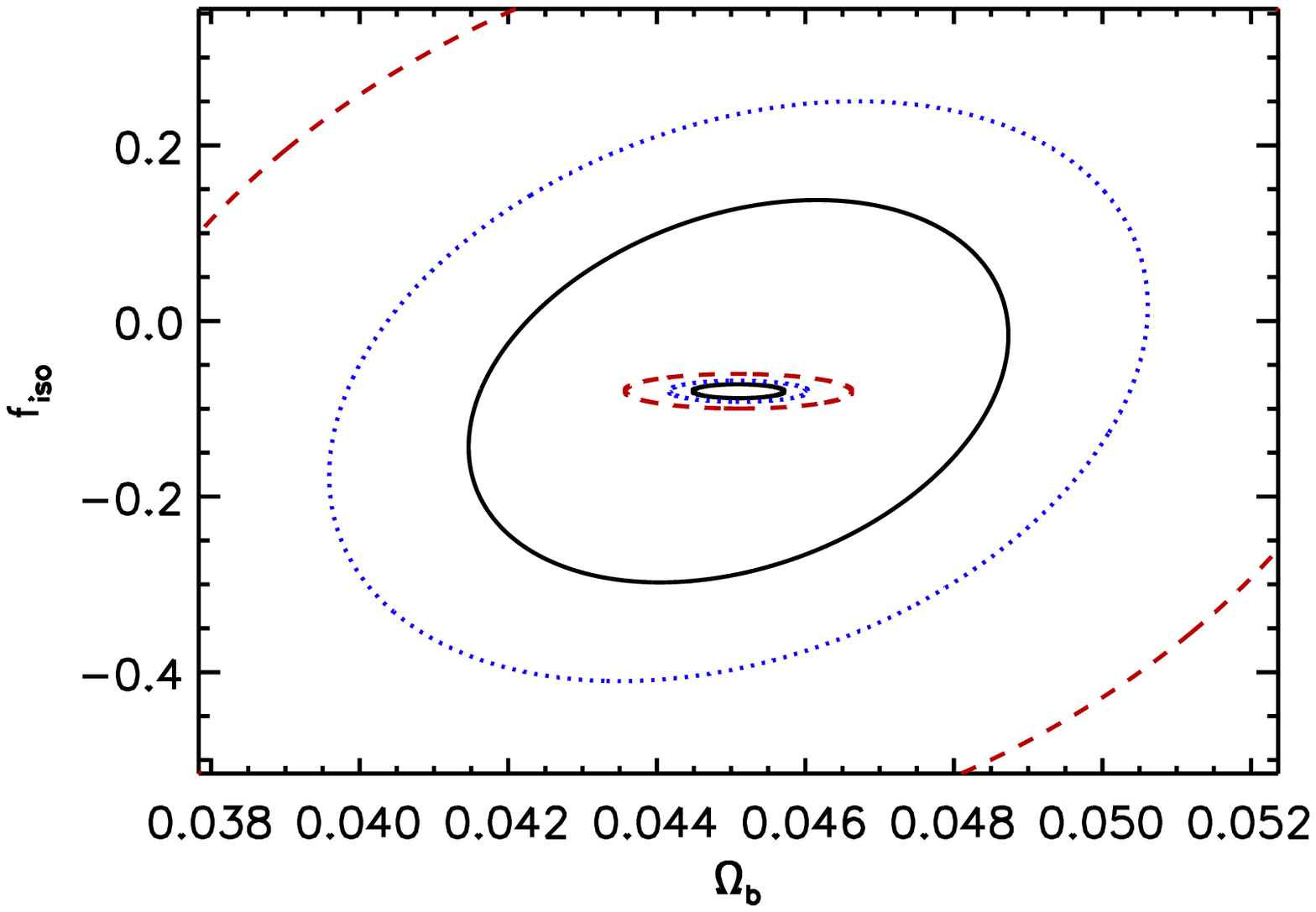} \\
\includegraphics[width=7.3cm]{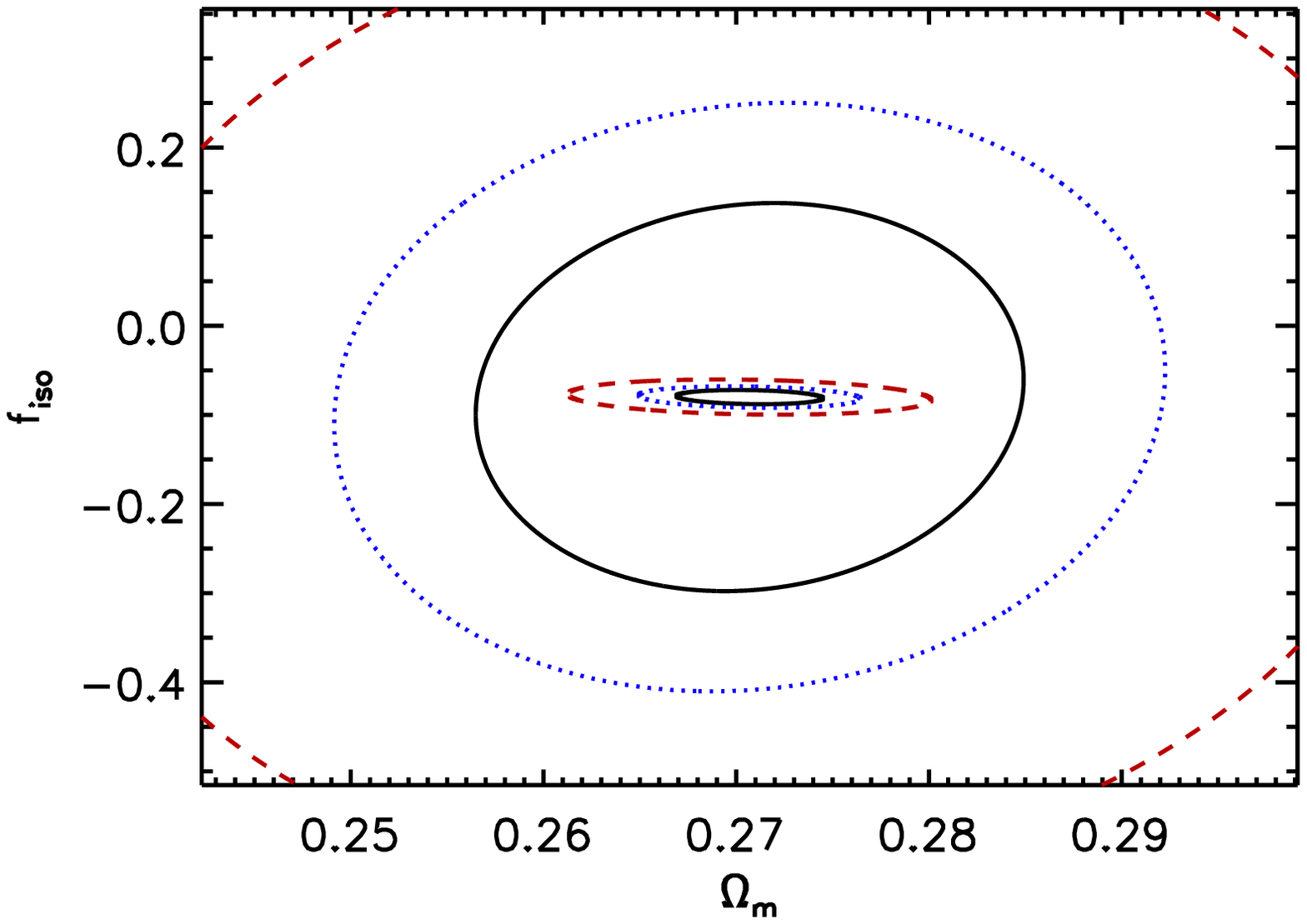}&
\includegraphics[width=7.3cm]{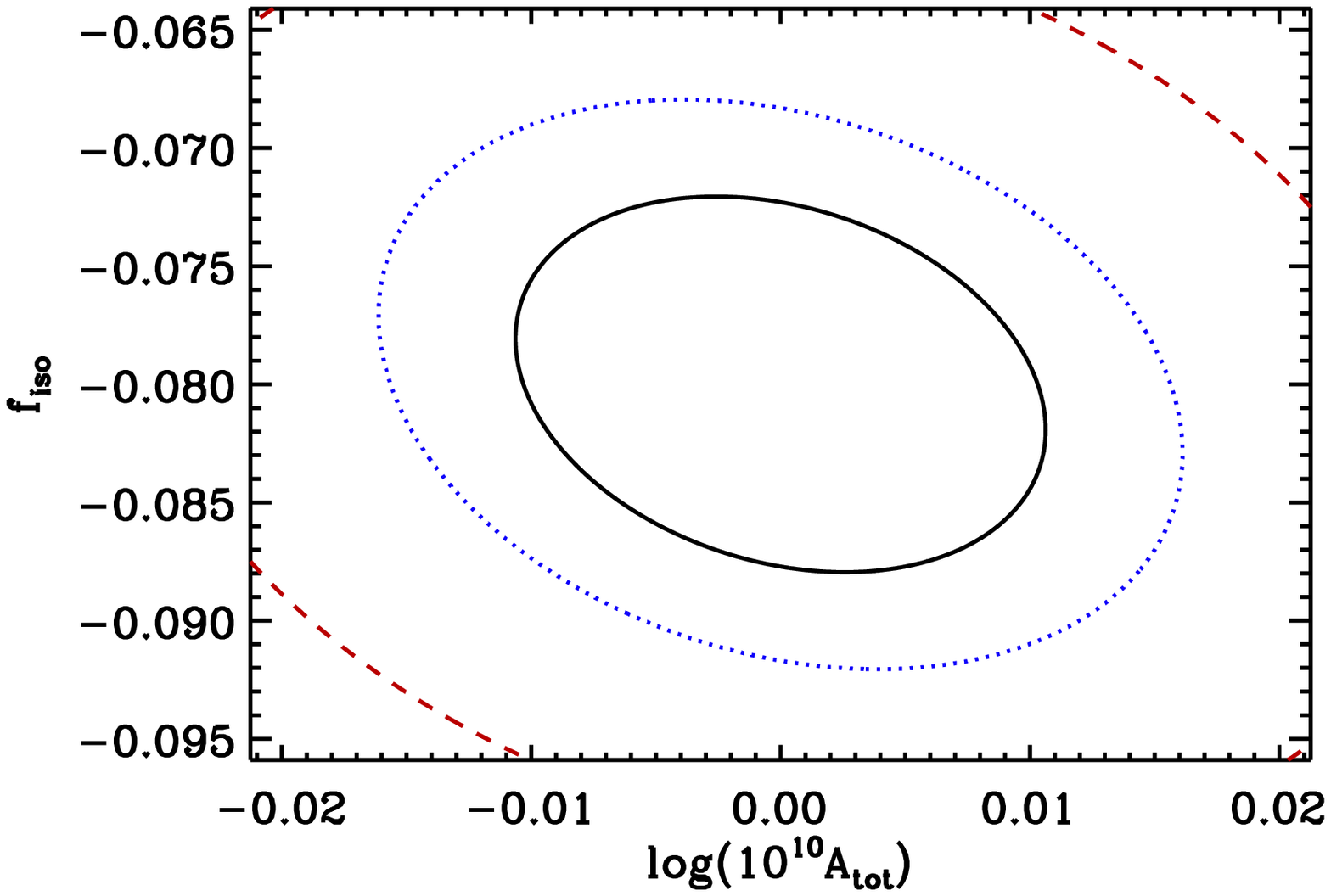}\\
\end{tabular}
\caption{2-parameter confidence levels for $f_{iso}$ and $q_\alpha$ with
  $q_\alpha= h, \Omega_b, \Omega_m, \log (10^{10}\,A_{\rm tot})$ for the fiducial model with
 $f_{iso}=-0.08$ obtained for an Euclid-like survey only (larger contours) and after combining the survey data with
  Planck. For $\log (10^{10} \,A_{\rm tot})$ we show the constraints from LSS+CMB only since for the LSS alone, using the $P(k)$-method, we marginalize over the power spectrum normalization. The blue dotted line and the red dashed line  represent the 68$\%$ C.L., 95.4$\%$
    C.L., respectively. The black solid line shows
    the 1-parameter confidence level at 1--$\sigma$. }
\label{fig_error2}
\end{figure*}
\begin{figure*}%[!ht]
\begin{tabular}{l l}
\centering
\includegraphics[width=7.3cm]{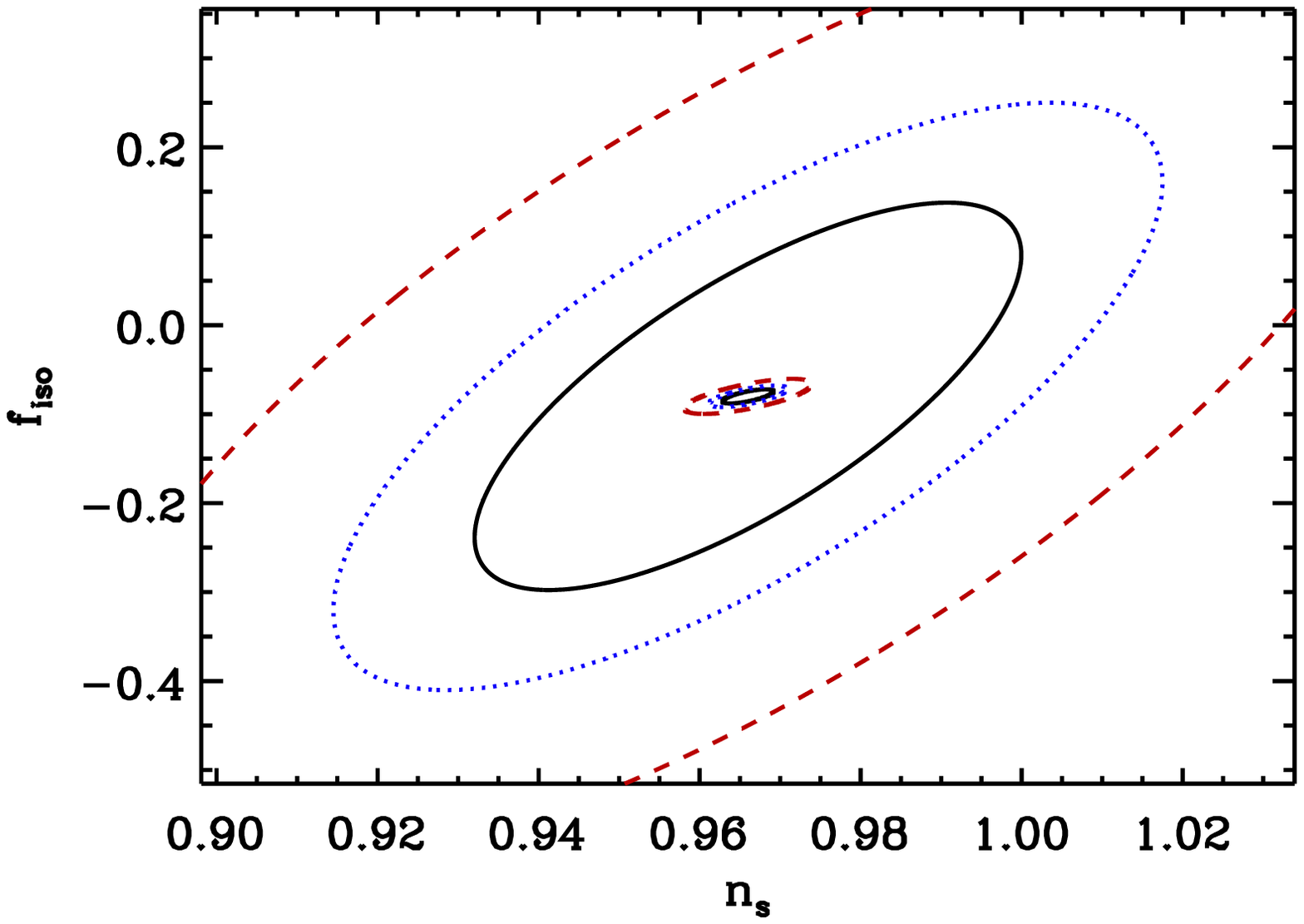}&
\includegraphics[width=7.3cm]{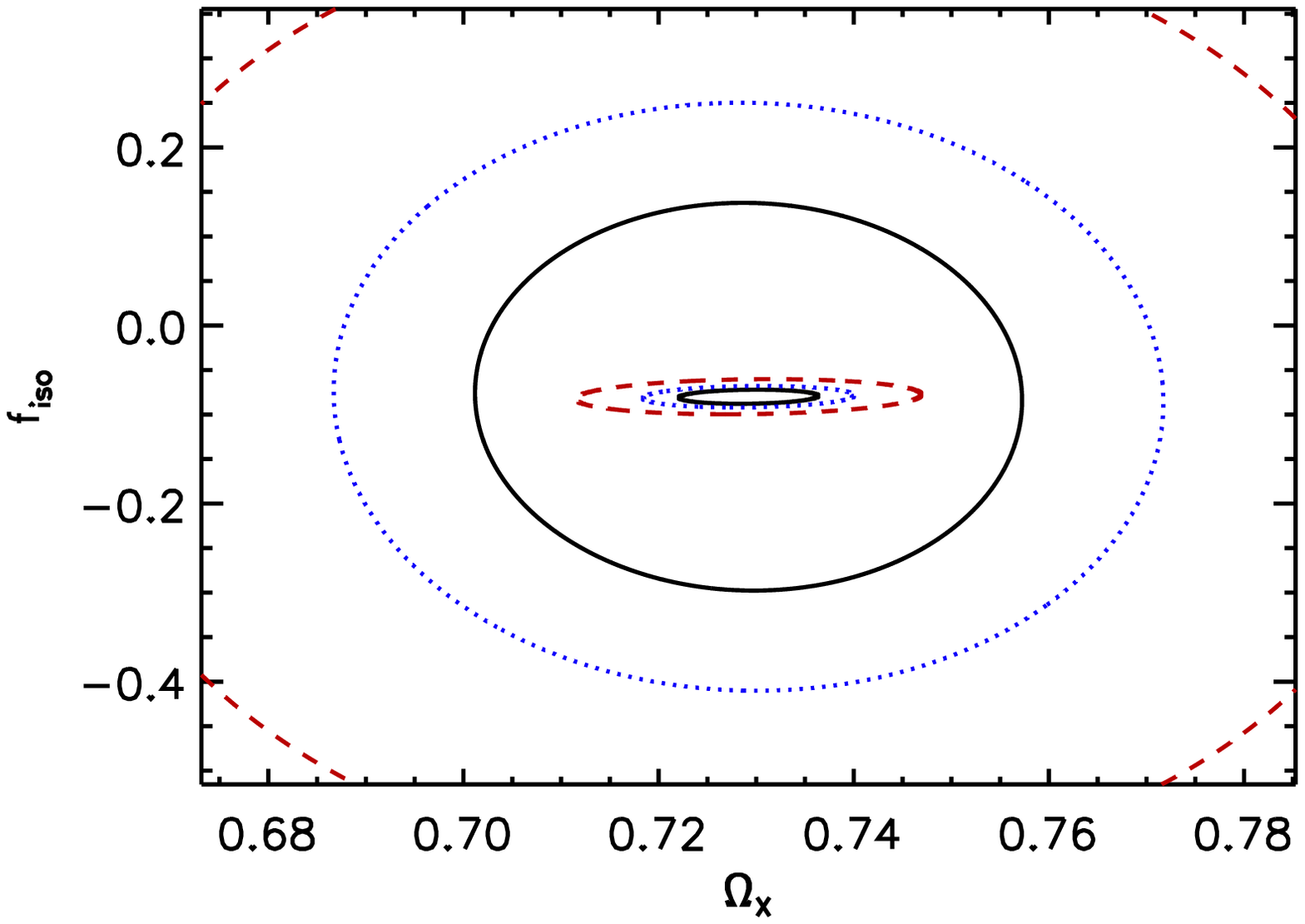} \\
\includegraphics[width=7.3cm]{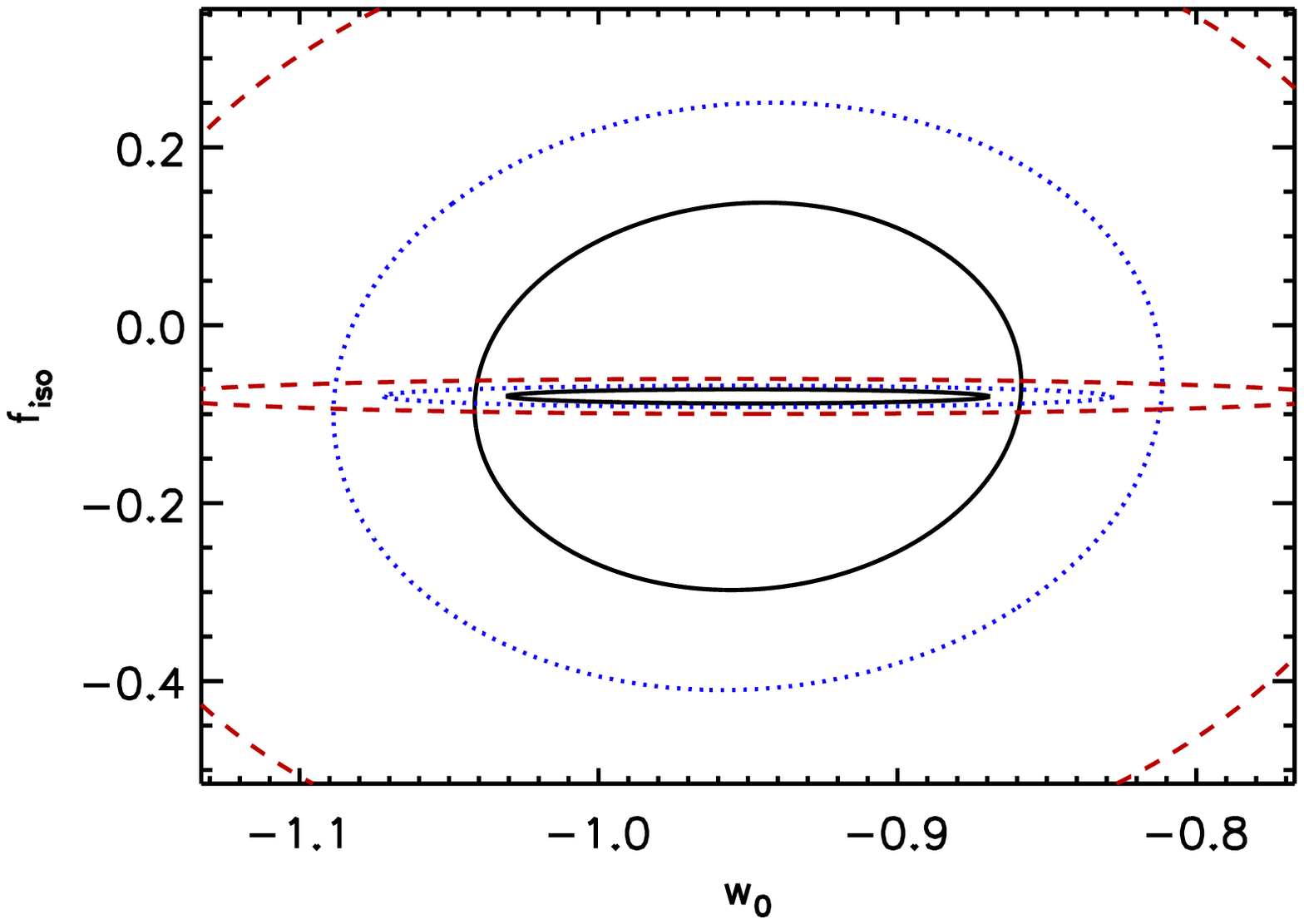}&
\includegraphics[width=7.3cm]{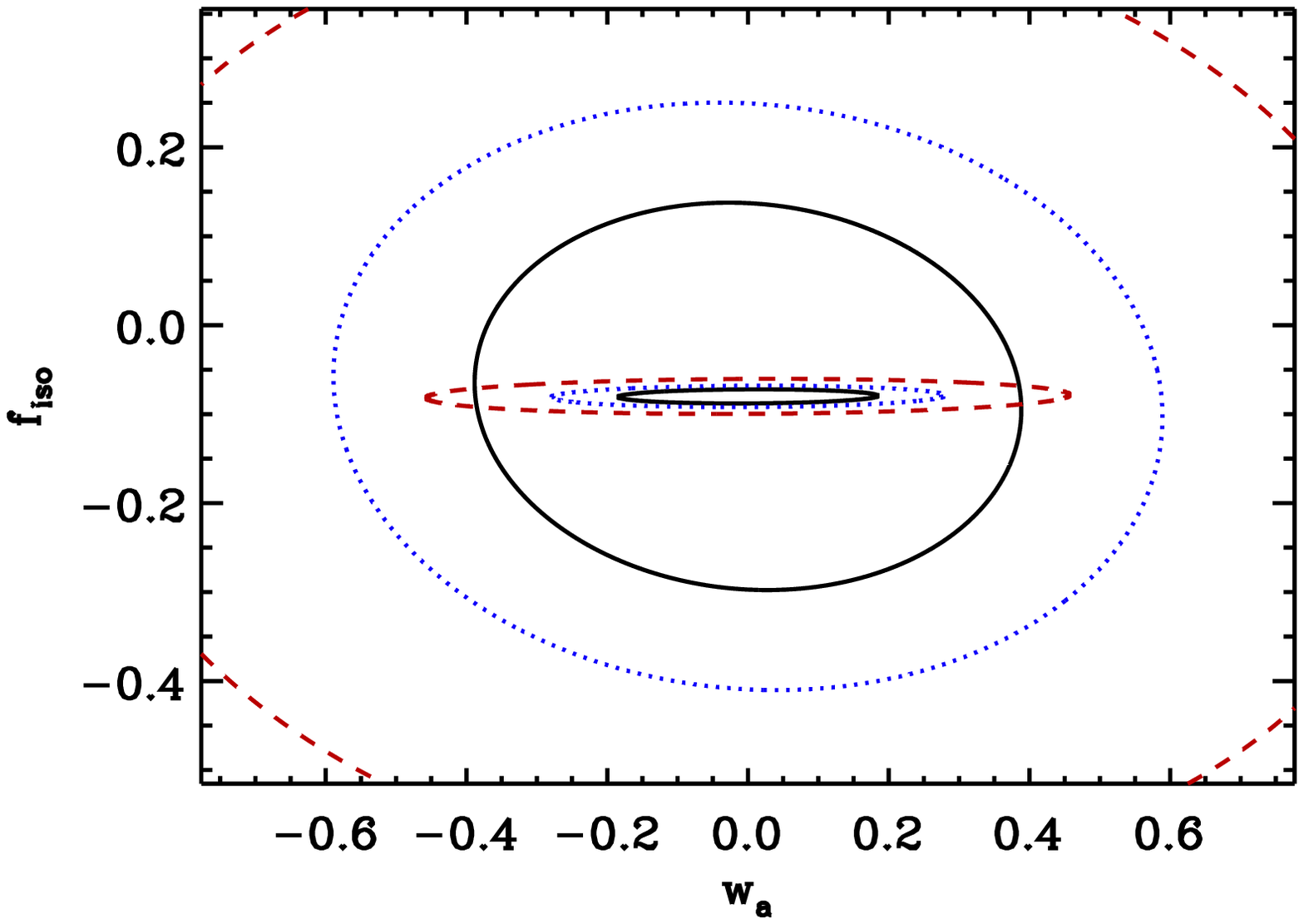}\\
\end{tabular}
\caption{2-parameter confidence levels for $f_{iso}$ and $q_\alpha$ with
  $q_\alpha=n_s,\Omega_{X0}, w_0,w_a$ for the fiducial model with
 $f_{iso}=-0.08$ obtained for an Euclid-like survey only (larger contours) and after combining the survey data with
  Planck. The blue dotted line and the red dashed line  represent the 68$\%$ C.L., 95.4$\%$
    C.L., respectively. The black solid line shows
    the 1-parameter confidence level at 1--$\sigma$. }
\label{fig_error}
\end{figure*}
 \begin{table*} 
%\centering
\begin{center}
\begin{tabular}{|ccccccc|}
\hline
\hline
\multicolumn{7}{c}{LSS+CMB}\\ 
%{}&{}&{}&LSS+CMB&{}&{}&{}\\
\hline
\footnotesize{$q_\alpha$}&\footnotesize{$\hat{q}_\alpha$}&\footnotesize{$\sigma(q_\alpha)$}&
\footnotesize{$r(q_\alpha$)}&\footnotesize{$\frac{\delta\theta_\alpha}{\delta
    f_{iso}}|_{\delta f_{iso}=1}$} & \footnotesize{$\frac{\delta\theta_\alpha}{\delta f_{iso}}|_{\delta f_{iso}=0.08}$} & 
\footnotesize{$\frac{\delta\theta_\alpha}{\delta f_{iso}}|_{\delta f_{iso}=0.01}$}\\
%&$|\frac{\delta\theta_\alpha}{\sigma_\alpha}|_{\delta f_{iso}=0.01}(\%)$\\
\hline
$ \Omega_{m_0}$&$0.2707 $&0.00379&-0.165 &$-0.0788$ & $-0.00630$ & $ -0.000788$  \\            
$\Omega_{X_0} $&$0.7293 $&0.00712&0.124&$ 0.1117 $ & $ 0.008939$ & $0.00111 $ \\
$ \Omega_{b_0}$&$0.0451 $&0.00061&-0.014 &$ -0.00101 $ & $ -8.793 \, 10^{-5}$ & $-1.01\, 10^{-5} $ \\ 
$h$&$0.703 $&0.0047&-0.054 &$-0.0321 $ & $ -0.00257$ & $ -0.000321$ \\
$n_s$&$0.966 $&0.00317 &0.69 &$  0.2770 $ & $ 0.02216 $ & $0.00277 $ \\                    
$w_0$&$-0.95 $&0.08349&-0.02  &$ -0.242 $ & $  -0.0193$ & $-0.00242 $\\                        
$w_a$&$0 $&0.18390&0.08  &$ 1.9786 $ & $ 0.1583$ & $ 0.019786$  \\
\footnotesize{$\log (10^{10} A_{\rm tot})$}&$1.3909 $&0.0106&-0.24  &$  -0.7471$ & $ -0.0597$ & $ -0.00747 $ \\                 
$f_{iso}$&$-0.08$& 0.00794&1&$1$   &$0.08$ &$0.01$\\
\hline
\hline
\hline
\multicolumn{7}{c}{LSS}\\
%{}&{}&{}&LSS&{}&{}&{}\\
\hline
$ \Omega_{m_0}$&$0.2707 $&0.01421&0.090&$ 0.00593$ & $0.000474915$ & $ 0.0000593 $ \\%& \\             
$\Omega_{X_0} $&$0.7293 $&0.02803&-0.018& $-0.00241$ & $-0.000193233$ & $-0.0024100$ \\%&152.7\\  
$ \Omega_{b_0}$&$0.0451 $& 0.00363&0.291&$ 0.00486 $ & $0.00038940 $ & $0.0000486 $ \\ %&43.24\\        
$h$&$0.703 $& 0.0123&0.246 &$0.01392 $ & $ 0.0011138 $ & $0.0001392 $ \\ %&0.25\\    
$n_s$&$0.966 $&0.0339&0.729&$0.11373   $ & $0.00909873  $ & $0.0011373 $ \\ %&3.93\\                        
$w_0$&$-0.95 $&0.0914&0.060& $0.02551  $ & $   0.00204113 $ & $0.0002551 $\\ %&120.55\\                           
$w_a$&$0 $&0.38839&-0.07 &$   -0.12471 $ & $  -0.00997681$ & $ -0.0012471 $ \\ %&123.0\\                                    
$f_{iso}$&$-0.08$&0.21771&1 &$1$ &$0.08$ &$0.01$ \\ %%&/\\
\hline
\end{tabular}
\caption{Forecast 1--$\sigma$ errors of the cosmological
  parameters $q_\alpha$ considered in the text and corresponding
  correlations with $f_{iso}$, for
an Euclid-like experiment alone and in combination with Planck. In
particular, in the 2${}^{\rm nd}$ column we report the fiducial
values adopted for the parameters listed in the 1${}^{\rm st}$
column. The corresponding 1--$\sigma$ errors are reported in the
3${}^{\rm rd}$ column, and the $q_\alpha$--$f_{iso}$ correlation coefficients in the
4${}^{\rm th}$ column. Finally, in the 5${}^{\rm th}$, 6${}^{\rm th}$
and 7${}^{\rm th}$ columns we list the shifts of the parameters
$\theta_\alpha$ for $\delta f_{iso}=1, 0.08,0.01$, respectively.}
\label{t:cmb+bao}
\end{center}
\end{table*}

In this work we consider a cosmological model
 with curvature, isocurvature and varying dark energy.
In this case, the combination of a Planck-like CMB experiment and a LSS probe
improves the constraints on the cosmological parameters by at least
one order of magnitude with respect to the Planck-only constraints for
a less general 
model with curvature and isocurvature but with a cosmological constant (see e.g.~\cite{Bucher00}).

Here, we have extended the analysis of Ref.~\cite{Mangilli10} to the
more general model described in \S \ref{sec:method}, and, moreover, 
we have added the cosmological information from the LSS to the CMB-only constraints.
The results are summarized in  the first three columns of Tab.~\ref{t:cmb+bao}.
Note that the 1-$\sigma$ error on $f_{iso}$ obtained from the joint
CMB+LSS analysis is $ \simeq 0.008$; this is slightly better than the
error expected from Planck alone, $ \sigma({f_{iso}})=0.01$ of
Ref.~\cite{Mangilli10}, where  $\Omega_k=0$ and $w=-1$ (both kept as fixed parameters) were assumed.

 In Figs.~\ref{fig_error2}-\ref{fig_error} we show the contours at 68\%
 CL, 95\% CL and 1-parameter CL at 1--$\sigma$, for $f_{iso}$ and $q_\alpha$ with
  $q_\alpha=\{n_s,\Omega_{X0}, w_0,w_a, h, \Omega_b, \Omega_m,
  \log (10^{10} \,A_{\rm tot})\}$, for the fiducial model with
 $f_{iso}=-0.08$, obtained from an Euclid-like survey alone (larger
 contours), and from the combination with Planck data. For
 $\log (10^{10} \,A_{\rm tot})$ we show only the constraints from LSS+CMB, since for the LSS alone, using the $P(k)$-method, we marginalize over the power spectrum normalization.  

 From Figs.~\ref{fig_error2}-\ref{fig_error} and  column 3 of Tab.~\ref{t:cmb+bao} it is evident that the CMB+LSS constraints improve dramatically with respect to the LSS-only case and also with respect to the CMB-only case \cite{Mangilli10,Bucher00}.
This means that adding LSS is a powerful tool which allows to break the strong degeneracies that affect CMB measurements when curvature, isocurvature and varying dark energy are included in the analysis.
As a consequence, constraints on the nature of the
initial conditions are greatly improved  by the combination CMB+LSS,
with respect to CMB alone. 

The next issue we address regards possible systematic errors on cosmological parameters due to incorrect assumptions about the nature of the initial conditions. 

To this purpose, we calculate the shifts on the cosmological parameters due to an incorrect  amount of the isocurvature fraction $f_{iso}$ in the fiducial model.

The results are summarized in the last three columns of
Tab.~\ref{t:cmb+bao}. As can be noticed, the absolute shifts (see
Eq.~(\ref{eq:shift})) $\delta\theta_\alpha/\delta f_{iso}$
assume the largest values for the dark energy parameters
$\Omega_{X_0}$, $w_0$, $w_a$, for the spectral index $n_s$, and for the
amplitude $A_{\rm tot}$ (see the $4^{th}$ column in Tab.~\ref{t:cmb+bao}). 

In interpreting these results one must bear in  mind that, rather than
the shifts themeselves, it is important to determine if and in which
cases these shifts on cosmological parameters are larger than the
corresponding statistical errors ($3^{rd}$ column in Tab.~\ref{t:cmb+bao}).

We have assumed a fiducial value $f_{iso}= -0.08$, so, if the true underlying model were adiabatic (for which the``true" value for the isocurvature fraction is $f_{iso}=0$),  
$f_{iso}$ would be fixed to an incorrect value by $\delta f_{iso}=+0.08$. In this case, the correspondent shifts on the cosmological parameters are shown in the $5^{th}$ column
 of Tab.~\ref{t:cmb+bao}.
  For $\Omega_{m_0}$, $\Omega_{X_0}$, $A_{\rm tot}$ and $n_s$ (this last
  being the most affected parameter) the shifts are bigger than the
  1--$\sigma$ statistical errors, meaning that the corresponding
  degeneracies with $f_{iso}$ are not completly solved, and a wrong
  assumption on the amount of isocurvature perturbations produces a bias in the estimation of the cosmological parameters even when adding LSS to CMB. 
On the other hand, the shift is slightly smaller but still comparable to the 1--$\sigma$ error for  $w_a$. For the case $\delta f_{iso}=+0.08$, 
adding the LSS information breaks the degeneracy $f_{iso}$-$h$ with
respect to the CMB-only case, even if the $h$-shift is still
comparable to the statistical error. For the other parameters the
degeneracies are reduced and the shifts are smaller but still of the
same order of magnitude as the statistical 1--$\sigma$ errors.
 
 On the other hand, if we consider the smaller value $\delta f_{iso}=+0.01$, %,  as done for the analysis in Chapter \ref{chap:ISO}, 
 the LSS information from an Euclid-like survey solves the
 degeneracies with $f_{iso}$ for all the cosmological parameters
 except $n_s$, as shown in the $6^{th}$ column of
 Tab.~\ref{t:cmb+bao}. This is a remarkable result given that the
 analysis is done including curvature and a varying dark energy equation of state.

 Since the statistical errors are not very sensitive to the fiducial
 $f_{iso}$ value, the case $\delta f_{iso}=+0.01$ 
is representative to quantify the difference in modeling with the pure adiabatic case and the mixed underlying model with a fiducial $f_{iso}=-0.01$.

As shown in Figs.~\ref{fig_error2}-\ref{fig_error}, for CMB+LSS the constraints on parameters which are degenerate with $f_{iso}$ improve considerably.
However, note that  the shifts on parameters as $\Omega_{X}$,
$\Omega_{m0}$, $w_0$ and $w_a$ 
are larger than in the LSS-only case, as shown in column 5 of Tab.~\ref{t:cmb+bao}. 

 This is confirmed by the behavior of the correlation coefficient $r$
 (see Eq.~(\ref{correlation})), shown in column 4 of
 Tab.~\ref{t:cmb+bao}. For instance, if we consider the correlation
 $f_{iso}$-$\Omega_{X_0}$, we can observe that $f_{iso}$ and
 $\Omega_{X_0}$ are negligibly correlated for LSS measurements, but
 they are tightly correlated for CMB measurements; when combining the two information, the LSS help to alleviate the degeneracy but a level of correlation still
 remains. 

 In some cases the correlation coefficient $r$, as well as the shifts,
 can even change sign when adding Planck to LSS. This change follows the dominant correlations between $f_{iso}$ and the cosmological parameters either from LSS or CMB.
 
 The parameter for which the degeneracy with the isocurvature
 fraction parameter $f_{iso}$ is not solved even when combining the CMB and LSS information is the spectral index $n_s$. 
We interpret this as due in part to the fact that in this paper we
assume a curvaton underlying model as a working example for which 
$n_s \equiv n_{ad}=n_{iso}$.
Therefore, $n_s$ encodes the information of both the spectral indices
which are correlated with $f_{iso}$ for both CMB and LSS. In fact,
the effect of a variation of $n_{iso}$ on the $C_l$ and the $P(k)$ would be different even for a fixed $f_{iso}$.
Conversely, the effect of changing $f_{iso}$ would be different in CMB
and LSS if $n_{ad}$ were not to be equal to $n_{iso}$. 
For the current data, the effect of changing $f_{iso}$ is stronger in CMB which thus ``dominates'' over LSS.
Therefore, the magnitude of the degeneracy $n_s-f_{iso}$ is related to our choice
$n_s\equiv n_{ad}=n_{iso}$.
When $n_{iso}=n_{ad}$ the information is compressed in one parameter which in
some way is ``double-degenerate'' with $f_{iso}$ and adding LSS helps but not
enough to fully solve the degeneracy.

These results confirm once again the great potential of forthcoming
large scale galaxy surveys.

\section{Discussion and conclusions}
In this work we have relied on the Fisher matrix approach to
forecasts, which is not well suited to quantify systematic
effects. The errors we have reported are statistical only, which are
optimistic in the sense that we assume that systematic errors are
vastly sub-dominant.  Let us therefore discuss sources of systematic
errors of major concern and their possible consequences on our
results: these are the effects of non-linearities and galaxy bias.  

In the  analysis presented here we have used the linear theory matter
power spectrum and a scale-independent  galaxy bias. Our choice of the maximum k-mode ensures that the effects of non-linearity are quite mild and that eventually, before an analysis is performed on real data,  they could be  accurately modeled. Having considered only the mildly non-linear regime ensures that non-linearities will not erase or reduce significantly the signal. Reconstruction techniques    that  are being developed \cite{reconstruction1,reconstruction2} and/or are already being applied to surveys \cite{reid}  offer promising prospects. Thus while it will  be mandatory to include non-linearities in the actual data analysis, the forecasted errors
are not made artificially smaller by using the linear matter power spectrum to compute our Fisher matrices.

Including smaller, more non-linear scales might help to improve the errors but would worsen the systematic effect of non-linearities and of a possible scale-dependent and/or non-linear bias. In this work we have made the simplifying assumption that bias is scale-independent but that the redshift dependence is not well known and so it is marginalised over. Bias is known to be scale-independent on large scales  but scale dependent for small scales and/or very massive halos. While halo bias can be modeled, the bias of galaxies hosted in dark matter halos can be more complicated especially at small scales. Here we have considered scales  where the so-called two halo term dominates and so complicated bias arising from the details of galaxy formation and   halo occupation distributions do not matter.
In addition we have checked that most of the information, useful to break degeneracies between the sound horizon at radiation drag and  isocurvature contribution, come mostly from very large scales (those constraining the matter -radiation equality). The BAO signal is localized on scales where scale-dependent bias is more likely to be an issue, but the BAO location is relatively  robust to this.   Our reported error-bars could have been made smaller by considering also information about the growth of structures (i.e., redshift space distortions and the normalization of the power spectrum), however we decided to, conservatively, marginalise over the amplitude to be less sensitive to possible assumptions about galaxy bias.

As found in Ref.~\cite{Mangilli10} for a Planck-like CMB experiment, even a tiny isocurvature fraction contribution, if not accounted for, can lead to an incorrect determination of the cosmological parameters even for a standard flat model with a cosmological constant. In this case however, when the isocurvature parameter $f_{iso}$ is added to the modelling, the errors increase but the cosmological parameters are recovered correctly. The error degradation from a CMB-only analysis is particularly severe for models where  spatial flatness is not imposed and curvature is left as a free parameter.

In this work we extend the analysis done in Ref.~\cite{Mangilli10}
and present the results from a joint analysis of a CMB
Planck-like 
experiment and a LSS EULID-like galaxy survey, for a cosmological model that includes an isocurvature fraction in the initial conditions, varying dark energy and curvature. 

As shown in Figs.~\ref{fig_error2}-\ref{fig_error} and in
Tab.~\ref{t:cmb+bao}, we find that adding the cosmological information 
coming from an Euclid-like LSS probe strongly reduces the degeneracies introduced by the presence of an isocurvature contribution with respect to a Planck-like CMB probe alone. 

When isocurvature is included in the analysis, we find that the most affected parameter is the spectral index $n_s$ for which the degeneracy with the isocurvature fraction parameter $f_{iso}$ is never fully solved even when combining CMB and LSS. 
 This is because in this paper we assume a curvaton underlying model as a working example for which $n_s \equiv n_{ad}=n_{iso}$. Therefore $n_s$ encodes the information of both the spectral indices which are correlated with $f_{iso}$ for both CMB and LSS. 
In this case, from the combination CMB+LSS, we find an absolute
$n_s$-shift$=0.277$ which is larger than the statistical 1--$\sigma$
error by two order of magnitude. For all the cosmological parameters
the correlations with $f_{iso}$ and the corresponding shifts are
reported in Tab.~\ref{t:cmb+bao}. Our findings are quantitative only
for the  $n_{ad}=n_{iso}$ case, 
but can hold qualitatively for other models with $n_{iso} \simeq 1$, as for example the axion-like isocurvature model.
 In this case, since $n_{ad}$ is close to scale invariant, we expect
 that the effect will be quantitatively very similar.

We investigate also the effect on the recovered cosmological
parameters  in the case of the presence of a small amount of
isocurvature  which is however ignored in the
analysis. Ref.~\cite{Mangilli10} found that even a tiny isocurvature
fraction, if neglected, could bias the determination of ``standard
rulers'' like the sound horizon $r_s(z_d)$ at radiation drag and
introduce systematic biases  in the interpretation of BAO observables
that are larger than the  statistical error-bars forecast for future
and forthcoming surveys.  Ref.~\cite{Mangilli10} considered a flat
Universe where the dark energy is a cosmological constant, but in a
more general cosmology the effect can be dramatically worst.

Here we find that adding LSS information to CMB priors
eliminates the bias (i.e. reduces it well below the statistical
errors) in the case of a small amount of isocurvature fraction
$f_{iso} \sim {\cal O} (10^{-2})$, even for a general cosmological
model with varying dark energy and curvature. The only parameter for
which the systematic shift remains still comparable to the 1--$\sigma$
error is the spectral index $n_s$.

These remarkable results are obtained  by better exploiting the
potential cosmological information of  future LSS surveys by adding to
BAO measurements also the power spectrum shape,
following the so-called $P(k)$-method marginalised over growth information.
In fact, the additional cosmological information in the power spectrum
helps  to break degeneracies among the cosmological parameters, and
thus indirectly improves the constraints on the isocurvature parameter
$f_{iso}$.

Considering that -conservatively- we have not included information on
the power spectrum amplitude and on the growth of cosmological structures 
(we marginalize over redshift space distortions parameters and LSS
power spectrum normalization), we conclude highlighting once again the 
synergy and complementarity  between  CMB and LSS and the great
potential of future galaxy surveys.

\acknowledgments
CC acknowledge the support from the Agenzia Spaziale Italiana (ASI-Uni Bologna-Astronomy Dept. Euclid-NIS 
I/039/10/0), and MIUR PRIN 2008 ÒDark energy and cosmology with large galaxy surveysÓ.
AM and LV acknowledge support of  MICINN grant AYA2008-03531.
LV is supported by FP7-IDEAS-Phys.LSS 240117.
Part of this work was carried out during a  visit of CC at ICC-UB, sponsored in part by a grant from vicerectorat d'Innovaci\'o i Transfer\`encia del Coneixement de la UB.
\appendix
\section{Planck Fisher matrix}
\label{sec:Planck}
In this work we use the Planck mission parameter
constraints as CMB priors, by estimating the cosmological parameter errors
via measurements of the temperature and polarization power
spectra. 
As CMB anisotropies, with the exception of the integrated
Sachs-Wolfe effect, are not able to constrain the equation of state of
dark energy $(w_0,w_a)$\footnote{On the contrary, using $(w_0,w_a)$ as
model parameters to compute the CMB Fisher matrix could artificially
break exiting degeneracies.}, we follow the prescription laid out by
DETF \cite{DETF}. %%%{albrecht09}.

We do not include any B-mode in our
forecasts and assume no tensor mode contribution
to the power spectra. 
We use the 100 GHz, 143 GHz,
and 217 GHz channels as science channels. These channels
have a beam of $\theta_{\rm fwhm}=9.5'$, $\theta_{\rm fwhm}=7.1'$, and
$\theta_{\rm fwhm}=5'$, respectively, 
and sensitivities of $\sigma_T= 2.5 \mu K/K$, $\sigma_T= 2.2 \mu K/K$,
$\sigma_T= 4.8 \mu K/K$ for temperature, and $\sigma_P = 4\mu K/K$,
$\sigma_P = 4.2\mu K/K$, $\sigma_P = 9.8\mu K/K$ for polarization,
respectively.  
We take $f_{\rm sky} = 0.80$ as the sky fraction in order to account for
galactic obstruction, and use a minimum $\ell$-mode $\ell_{\rm
  min}=30$ in order to avoid problems with polarization foregrounds
and not to include information from the late
Integrated Sachs-Wolfe effect, which depends on the specific
dark energy model. 
We discard temperature and polarization data at
$\ell > 2000$  to reduce sensitivity to contributions from patchy
reionization and point source contamination (see \cite{DETF} and
references therein).

We assume a fiducial cosmology with an anti-correlated isocurvature contribution, varying dark energy and curvature.
Therefore we choose the
following set of parameters to describe the temperature and
polarization power spectra 
$\vec{\theta}= (\omega_m, \omega_b, f_{iso},
100\times \theta_S,\log (10^{10}A_{\rm tot}), n_S, w_0,w_a)$, where $\theta_S$ is the
angular size of the sound horizon at last scattering and $w_0$ and
$w_a$ are the dark energy parameters according to the CPL
parametrization of the dark energy equation of state $w(z) = w_0 + w_a\,z/(1+z)$.

The Fisher matrix for CMB power spectrum is given by \cite{Zalda:1997,Zalda:1997b}:
\begin{equation}
  F_{ij}^{CMB-Planck}=\sum_{l}\sum_{X,Y}\frac{\partial
    C_{X,l}}{\partial\theta_{i}}\mathrm{COV^{-1}_{XY}}\frac{\partial
    C_{Y,l}}{\partial\theta_{j}},
  \label{eqn:cmbfisher}
\end{equation}
where $\theta_i$ are the parameters to constrain, 
$C_{X,l}$ is the harmonic power spectrum for the
temperature-temperature ($X\equiv TT$), 
temperature-E-polarization ($X\equiv TE$) and the 
E-polarization-E-polarization ($X\equiv EE$) power spectrum. The covariance
$\rm{COV}^{-1}_{XY}$ of the errors for the various power spectra is
given by the fourth moment of the distribution, 
which under Gaussian assumptions is entirely given in terms of the $C_{X,l}$ with 
\begin{eqnarray}
{\rm COV}_{T,T} & = & f_\ell\left(C_{T,l}+W_T^{-1}B_l^{-2}\right)^2 \\
{\rm COV}_{E,E} & = & f_\ell\left(C_{E,l}+W_P^{-1}B_l^{-2}\right)^2  \\
{\rm COV}_{TE,TE} & = & f_\ell\Big[C_{TE,l}^2+\left(C_{T,l}+W_T^{-1}B_l^{-2}\right)\left(C_{E,l}+W_P^{-1}B_l^{-2}\right)\Big]\\
{\rm COV}_{T,E} & = & f_\ell C_{TE,l}^2  \\
{\rm COV}_{T,TE} & = & f_\ell C_{TE,l}\left(C_{T,l}+W_T^{-1}B_l^{-2}\right) \\
{\rm COV}_{E,TE} & = & f_\ell C_{TE,l}\left(C_{E,l}+W_P^{-1}B_l^{-2}\right)\; ,
\end{eqnarray}
where $f_\ell = 2/((2\ell+1)f_{\rm sky})$,
$W_{T,P}=\sum_c W^c_{T,P}$, $W^c_{T,P}=(\sigma^c_{T,P}\theta^c_{\rm fwhm})^{-2}$  
being the weight per solid angle for temperature and polarization respectively, 
with a 1--$\sigma$ sensitivity per pixel of $\sigma^c_{T,P}$ and a beam
of $\theta^c_{\rm fwhm}$ extent, for each frequency channel $c$. 
The beam window function is given in terms of the full width half
maximum (fwhm) beam width by 
$B_{\ell}^2 =\sum_c (B^c_{\ell})^2 W^c_{T,P}/W_{T,P}$, where 
$(B^c_\ell)^2= \exp\left(-\ell(\ell+1)/(l^c_s)^2\right)$, 
$l^c_s=(\theta^c_{\rm fwhm})^{-1}\sqrt(8\ln2)$ 
and $f_{\rm sky}$ is the sky fraction \cite{9702100}. 

We then calculate the Planck CMB Fisher matrix with the help of the
publicly available CAMB code \cite{Lewis:1999bs}. Finally, we transform 
the Planck Fisher matrix for the DETF parameter set to the
final parameter sets $\bf q$ considered in this work (see \S \ref{sec:method}),
using the transformation
\begin{equation}
F_{\alpha \beta}^{CMB-Planck}= \sum_{ij} \frac{\partial
  \theta_i}{\partial q_{\alpha}}\,                       
F_{ij}^{CMB-Planck}\, \frac{\partial \theta_j}{\partial q_{\beta}}.
\end{equation}


\begin{thebibliography}{99}

\bibitem{mukhanov-adiabatic}
V.~ Mukhanov, H.A. ~Feldman and R.H.~ Brandenberger,
\newblock Phys. Rept. {\bf 15}, 203 (1992).


\bibitem{Brandenberger92}
Brandenberger, R., Mukhanov, V., \& Prokopec, T.\
1992, Physical Review Letters, 69, 3606

\bibitem{KomatsuWMAP7} 
Komatsu, E., et al.\ 
2010, \arXivid{1001.4538}

\bibitem{Linde1985}
A.~D.~Linde,
%``Generation Of Isothermal Density Perturbations In The Inflationary Universe,''
Phys.\ Lett.\ B {\bf 158}, 375 (1985);
A.~D.~Linde and V.~Mukhanov,
%``Nongaussian isocurvature perturbations from inflation,''
Phys.\ Rev.\ D {\bf 56}, 535 (1997);
\bibitem{Kofman1986}
L.~A.~Kofman and A.~D.~Linde,
%``Generation Of Density Perturbations In The Inflationary Cosmology,''
Nucl.\ Phys.\ B {\bf 282}, 555 (1987);

\bibitem{Mollerach1990}
S.~Mollerach,
%``On The Primordial Origin Of Isocurvature Perturbations,''
Phys.\ Lett.\ B {\bf 242}, 158 (1990);


\bibitem{Polarski1994}
D.~Polarski and A.~A.~Starobinsky,
%``Isocurvature perturbations in multiple inflationary models,''
Phys.\ Rev.\ D {\bf 50}, 6123 (1994);
%\bibitem{Sasaki:1995aw}
M.~Sasaki and E.~D.~Stewart,
%``A General analytic formula for the spectral index of the density perturbations produced during inflation,''
Prog.\ Theor.\ Phys.\  {\bf 95}, 71 (1996);
%\bibitem{Sasaki:1998ug}
M.~Sasaki and T.~Tanaka,
%``Super-horizon scale dynamics of multi-scalar inflation,''
Prog.\ Theor.\ Phys.\  {\bf 99}, 763 (1998).

\bibitem{Langlois99}
 Langlois, D.,
   % title = "{Correlated adiabatic and isocurvature perturbations from double inflation}",
   PhysRevD.59, 123512 (1999)
  \bibitem{Peebles1999}
P.~J.~E.~Peebles,
%``An Isocurvature Cold Dark Matter Cosmogony. I. A Worked Example of Evolution through Inflation,''
Astrophys.\ J. {\bf 510}, 523 (1999).

\bibitem{Bartolo2001}
N.~Bartolo, S.~Matarrese and A.~Riotto,
%``Adiabatic and isocurvature perturbations from inflation: Power spectra  and consistency relations,''
Phys.\ Rev.\ D {\bf 64}, 123504 (2001); 
C.~Gordon, D.~Wands, B.~A.~Bassett and R.~Maartens,
%``Adiabatic and entropy perturbations from inflation,''
Phys.\ Rev.\ D {\bf 63}, 023506 (2001);
%\bibitem{Wands2002}
D.~Wands, N.~Bartolo, S.~Matarrese and A.~Riotto,
%``An observational test of two-field inflation,''
Phys.\ Rev.\ D {\bf 66}, 043520 (2002).
%\bibitem{Garcia-Bellido1995}
J.~Garc\'{\i}a-Bellido and D.~Wands,
%``Metric perturbations in two-field inflation,''
Phys.\ Rev.\ D {\bf 53}, 5437 (1996); {\bf 52}, 6739 (1995).
%[astro-ph/9511029].

%%Bucher general (+neutrinos iso)

\bibitem{Bucher-general-2001}
M.~Bucher\&al. 
\newblock Phys. Rev. D {\bf 62}, 083508 (2000)

%%%axion

\bibitem{Axenides:1983hj}
  M.~Axenides, R.~H.~Brandenberger and M.~S.~Turner,
  %``Development Of Axion Perturbations In An Axion Dominated Universe,''
  Phys.\ Lett.\ B {\bf 126} (1983) 178.
  %%CITATION = PHLTA,B126,178;%%


\bibitem{Lindeaxion} 
  A.~D.~Linde, 
  %``Generation Of Isothermal Density Perturbations In The Inflationary 
  %Universe,'' 
  JETP Lett.\  {\bf 40} (1984) 1333 
  [Pisma Zh.\ Eksp.\ Teor.\ Fiz.\  {\bf 40} (1984) 496]; 
  %%CITATION = JTPLA,40,1333;%% 
  %A.~D.~Linde, 
  %``Generation Of Isothermal Density Perturbations In The Inflationary 
  %Universe,'' 
  Phys.\ Lett.\ B {\bf 158}, 375 (1985); 
  %%CITATION = PHLTA,B158,375;%% 
  %A.~D.~Linde, 
  %``Inflation And Axion Cosmology,'' 
  Phys.\ Lett.\ B {\bf 201} (1988) 437. 
  %%CITATION = PHLTA,B201,437;%% 
 
\bibitem{SeckelTurner}
  D.~Seckel and M.~S.~Turner,
  %``'Isothermal' Density Perturbations In An Axion Dominated Inflationary
  %Universe,''
  Phys.\ Rev.\ D {\bf 32} (1985) 3178. 
  %%CITATION = PHRVA,D32,3178;%%

\bibitem{hybrid}
  A.~D.~Linde, 
  %``Axions in inflationary cosmology,''
  Phys.\ Lett.\ B {\bf 259} (1991) 38.
  %%CITATION = PHLTA,B259,38;%% 
   
\bibitem{TurnerWilczek} 
  M.~S.~Turner and F.~Wilczek, 
  %``Inflationary Axion Cosmology,'' 
  Phys.\ Rev.\ Lett.\  {\bf 66} (1991) 5. 
  %%CITATION = PRLTA,66,5;%%

\bibitem{LindeLyth}
 A.~D.~Linde and D.~H.~Lyth,
 Phys.\ Lett.\ B {\bf 246} (1990) 353.

\bibitem{Lyth:1991}
  D.~H.~Lyth,
  %``Axions And Inflation: Sitting In The Vacuum,''
  Phys.\ Rev.\ D {\bf 45} (1992) 3394.
  %%CITATION = PHRVA,D45,3394;%%
\bibitem{Shellard:1997}
  E.~P.~S.~Shellard and R.~A.~Battye,
  %``Cosmic axions,''
  \arXivid{astro-ph/9802216}.
  
  %\cite{Kawasaki:1995ta}
\bibitem{Kawasaki1995axion}
M.~Kawasaki, N.~Sugiyama and T.~Yanagida,
%``Isocurvature and Adiabatic Fluctuations of Axion in Chaotic Inflation Models and Large Scale Structure,''
Phys.\ Rev.\ D {\bf 54}, 2442 (1996);
%[hep-ph/9512368].

%curvaton
\bibitem{curvaton1}
A. D. ~ Linde and V. F. ~ Mukhanov,
\newblock Phys. Rev. D{\bf 56}, 535 (1997), \arXivid{astro-ph/9610219}

\bibitem{curvaton2}
D.~H. Lyth and  D.~Wands,
\newblock Phys. Lett.B {\bf 524}, 5-14 (2002).

\bibitem{curvaton3}
D.~H. Lyth, C.~Ungarelli, and D.~Wands,
\newblock Physical Review D {\bf 67}, 023503 (2003).
% check K.  ~Enqvist and M. S. ~ Sloth, 
%\newblock Nucl. Phys. {\bf B 626}, 395 (2002);


\bibitem{Efstathiou86}
G.~Efstathiou and J.R.~Bond,
``Isocurvature cold dark matter fluctuations''
\newblock MNRS {\bf 218}, 103 (1986)

\bibitem{Enqvist02}
K.~Enqvist, H.~Kurki-Suonio and J.~Valiviita,
\newblock  Phys. Rev. D {\bf 65}, 043002 (2002)

%pure iso model in curved background also ruled out
%%Wmap 1st peak position
\bibitem{Page03}
L.~Page et al.
\newblock Astrophys. J. Suppl. {\bf 148} 233 (2003) 


\bibitem{Hinshaw06}
G.~Hinshaw et al.
\newblock Astrophys. J. Suppl.{\bf 170} 288 (2007)



%%%%papers Maria Beltran
\bibitem{Beltran-04}
Maria Beltran, Juan Garcia-Bellido, Julien Lesgourgues, Alain Riazuelo
Journal-ref: Phys.Rev. D70 (2004) 103530 

\bibitem{Beltran-05}
Maria Beltran, Juan Garcia-Bellido, Julien Lesgourgues, Matteo Viel
Phys.Rev.D72:103515,2005 
%%%%%

\bibitem{Valiviita09}
J.~Valiviita and T.~Giannantonio,
\newblock  Phys. Rev. D {\bf 80}, 123516 (2009)

\bibitem{Dunkley05}
J.~Dunkley et al.,
Phys. Rev. Lett.  {\bf 95}, 261303 (2005)

%"Hints of isocurvature perturbations in the cosmic microwave background?"
\bibitem{Keskitalo07} 
Keskitalo, R., Kurki-Suonio, H., Muhonen, V., Valiviita, 
J.\ 2007, Journal of Cosmology and Astro-Particle Physics, 9, 8

\bibitem{Valiviita03}
J.~Valiviita and V.~Muhonen,
\newblock  Phys. Rev. Lett. {\bf 91}, 131302 (2003)

\bibitem{trotta2003} 
R.~Trotta
New Astron. Rev. {\bf 47}, 769-774 (2003)


\bibitem{Langlois00}
D.~Langlois and A.~Riazuelo
\newblock Phys. Rev. D {\bf 62}, 043504 (2000)

\bibitem{Bucher00}
M.~Bucher, K.~Moodley and N.~Turok,
Cosmology and Particle Physics, {\bf 555}, 313 (2001)

%"Cold dark matter isocurvature perturbations: Constraints and model selection"
\bibitem{Sollom09} 
Sollom, I., Challinor,A., \& Hobson, M.~P.\ 2009, \prd, 79, 123521

%%cmb polarization and iso + general 
\bibitem{bucher-pol}
M.~Bucher, K.~Moodley and N.~Turok,
\newblock Physical Review Lett {\bf 87}, 191301 (2001).

\bibitem{Mangilli10}A. Mangilli, L.Verde and M.Beltran,
\newblock JCAP {\bf 10} 009 (2010), \arXivid{1006.3806}

\bibitem{Kurki05}
H.Kurki-Suonio, V.~Muhonen and J.~Valiviita
\newblock Phys. Rev. D {\bf 71}, 063005 (2005)

\bibitem{Melita11}
C. Carbone, L. Verde, Y. Wang and A. Cimatti,
JCAP{\bf 03} 030 (2011)



%%%iso general theory papers
%%general iso - compensation + notation
\bibitem{Hu-Sper-Whi97}
W.~Hu, D.N.~Spergel and M.~White,
\newblock Phys.Rev. D {\bf 55}, 3288-3302 (1997)

%%Bucher general (+neutrinos iso)

%fiso
\bibitem{Peiris03}
H.V.~Peiris et al.,
``First-Year Wilkinson Microwave Anisotropy Probe (WMAP) Observations: Implications For Inflation''
\newblock Astophys. J. Suppl. {\bf 148}, 213 (2003)


\bibitem{WMAP5}
Komatsu, E., et al. 2009, \emph{Astrophys. J. Suppl.}, 180, 330



\bibitem{Larson11}
D.~Larson
\newblock Astrophys.J.Suppl. {\bf 192} 16 (2011)


%curvaton
%\bibitem{curvaton1}
%A. D. ~ Linde and V. F. ~ Mukhanov,
%\newblock Phys. Rev. D{\bf 56}, 535 (1997), astro-ph/9610219

%\bibitem{curvaton2}
%D.~H. Lyth and  D.~Wands,
%\newblock Phys. Lett.B {\bf 524}, 5-14 (2002).

%\bibitem{curvaton3}
%D.~H. Lyth, C.~Ungarelli, and D.~Wands,
%\newblock Physical Review D {\bf 67}, 023503 (2003).
% check K.  ~Enqvist and M. S. ~ Sloth, 
%\newblock Nucl. Phys. {\bf B 626}, 395 (2002);

%Fisher
\bibitem{Fisher1935}
R.A.~Fisher, 
\newblock J. Roy. Stat. Soc., {\bf 98}, 39 (1935)

\bibitem{YB:0912.0914}
R. Laureijs, et al. (Euclid Science Study Team), \arXivid{0912.0914}

\bibitem{FKP}
Feldman, H., A., Kaiser, N., \& Peacock, J., A. 1994,
\emph{Astrophys. J.}, 426, 23

%%CPL DE parametrization
\bibitem{CPL01}
M. Chevallier and D. Polarski, Int. J. Mod. Phys. D 10, {\bf 213} (2001).

\bibitem{CPL03}
E. V. Linder, 
Phys. Rev. Lett. {\bf 90}, 091301 (2003).

\bibitem{VS96}
Vogeley, M. S. \& Szalay, A. S. 1996
\emph{Astrophys. J.}, 465, 43

\bibitem{Tegmark}
Tegmark, M., Taylor A., Heavens A. 1997, \emph{Astrophys. J.}, 480, 22

\bibitem{Jungman}
Jungman, G., Kamionkowski, M., Kosowsky, A., Spergel,
D. 1996, \emph{Phys. Rev. D}, 54, 1332

\bibitem{SE03}
Seo, H., J., \& Eisenstein, D., J. 2003,
\emph{Astrophys. J.}, 598, 720


\bibitem{EisensteinHu} Eisenstein, D.~J., \& Hu, W.\ 1998, ApJ, 496, 605 

\bibitem{DETF} Albrecht, A., et al.\ 
2006, \arXivid{astro-ph/0609591} 

\bibitem{9605017}
W. E. Ballinger, J. A. Peacock, and A. F. Heavens, Mon.Not.Roy.As.Soc. 282, 877 (1996), a

\bibitem{Seljak00}
Seljak, U. 2000, \emph{Mont. Not. Roy. Astron. Soc.}, 318, 203

\bibitem{Kaiser1987}
Kaiser, N. 1987, \emph{Mont. Not. Roy. Astron. Soc.}, 227, 1 

\bibitem{astro-ph/0305286}
Linder, E. V. \& Jenkins, A. 2003,
\emph{Mont. Not. Roy. Astron. Soc.}, 346, 573

\bibitem{CAMB} Lewis, A., Challinor, A., \& Lasenby, A.\ 2000,
  \emph{Astrophys. J.}, 538, 473

\bibitem{0904.2218}
Wang, Y. 2010, \emph{Mod. Phys. Lett. A}, 25, 3093

\bibitem{0710.3885}
Wang, Y. 2008, \emph{Journal of Cosmology and Astro-Particle Physics},
05, 021 

\bibitem{Tegmark97}Tegmark, S. 1997,
\emph{Phys. Rev. Lett.}, 79, 3806

\bibitem{Wang06}
Wang, Y. 2006, \emph{Astrophys. J.}, 647, 1

\bibitem{Heavens-Kitching-Verde07}
A. F.~Heavens, T.D.~Kitching and L.~Verde
\newblock Mon. Not. Roy. Astron. Soc. {\bf 380},1029-1035 (2007)

\bibitem{Wang08a}
Wang, Y. 2008, \emph{Phys. Rev. D}, 77, 123525

\bibitem{1006.3517}
Wang, Y., et al. 2010, \emph{Mont. Not. Roy. Astron. Soc.}, 409, 737

\bibitem{Geach10}
Geach, J. E.; Cimatti, A.; Percival, W.; Wang, Y.; Guzzo, L.; 
Zamorani, G.; Rosati, P.; Pozzetti, L.; Orsi, A.; Baugh, C. M.; 
Lacey, C. G.; Garilli, B.; Franzetti, P.; Walsh, J. R.; Kümmel, M.,
2010, \emph{Mont. Not. Roy. Astron. Soc.}, 402, 1330
%arXiv:0911.0686

\bibitem{Orsi10}
Orsi, Alvaro; Baugh, C. M.; Lacey, C. G.; Cimatti, A.; Wang, Y.;
Zamorani, G.,
\arXivid{0911.0669}, \emph{Mont. Not. Roy. Astron. Soc.}, {\bf 402}
1330G (2010)


\bibitem{reconstruction1} Eisenstein, D.~J., Seo, H.-J., Sirko, E., \& Spergel, D.~N.\ 2007, ApJ, 664, 675

\bibitem{reconstruction2} Wagner, C., M{\"u}ller, V., \& Steinmetz, M.\ 2008, A\&A, 487, 63 


\bibitem{reid} Reid, B.~A., et al.\ 2010,
MNRAS, 404, 60

\bibitem{Zalda:1997}
Zaldarriaga M., Seljak U., 1997, PRD, 55, 1830

\bibitem{Zalda:1997b}
Zaldarriaga M., Spergel D. N., Seljak U., 1997, APJ, 488, 1

\bibitem{9702100}
Bond, J. R., Efstathiou, G., \& Tegmark, M. 1997
\emph{Mont. Not. Roy. Astron. Soc.}, 291, L33

\bibitem{Lewis:1999bs}
Lewis, A., Challinor, A., \& Lasenby, A. 2000,
\emph{Astrophys. J.}, 538, 473

\end{thebibliography}
\end{document}